 \documentclass{aa}
 \usepackage{psfig}

\newcommand{\more}{\raisebox{-1.1mm}{$\stackrel{>}{\sim}$}}
\newcommand{\msol}{{M$_{\odot}$}}

\newcommand{\HP}{{\sc hipparcos}}
\newcommand{\C}{Cepheids}

\begin{document}

\thesaurus{04(08.04.1, 08.22.1, 11.13.1, 12.04.3)}
\title{Multi-colour $PL$-relations of  Cepheids in the 
{\sc hipparcos} catalogue and the  distance to the LMC
\thanks{Based on data from the ESA \HP\ astrometry satellite.}
}

\author{M.A.T. Groenewegen \inst{1} \and R.D. Oudmaijer \inst{2,3} 
}

\offprints{Martin Groenewegen (groen@mpa-garching.mpg.de)}

\institute{
Max-Planck Institut f\"ur Astrophysik, Karl-Schwarzschild-Stra{\ss}e 1, 
D-85740 Garching, Germany
\and
Department of Physics and Astronomy, University of Leeds, LS2 9JT Leeds, U.K.
\and
Blackett Laboratory, Imperial College of Science, 
Technology and Medicine, Prince Consort Road, London SW7 2BZ, U.K.
}

\date{received: Nov. 22nd, 1999,  accepted: Feb. 3rd, 2000}

\authorrunning{Groenewegen \& Oudmaijer}
\titlerunning{The distance to the LMC from \HP\ Cepheid parallaxes}

\maketitle
 
\begin{abstract}

We analyse a sample of 236 \C\ from the \HP\ catalog, using the method
of ``reduced parallaxes'' in $V, I, K$ and the reddening-free
``Wesenheit-index''. We compare our sample to those considered by
Feast \& Catchpole (1997) and Lanoix et al. (1999), and argue that our
sample is the most carefully selected one with respect to
completeness, the flagging of overtone pulsators, and the removal of
\C\ that may influence the analyses for various reasons (double-mode
\C, unreliable \HP\ solutions, possible contaminated photometry due to
binary companions).

From numerical simulations, and confirmed by the observed parallax
distribution, we derive a (vertical) scale height of \C\ of 70 pc, as
expected for a population of 3-10 \msol\ stars. This has consequences
for Malmquist- and Lutz-Kelker (Lutz \& Kelker 1973, Oudmaijer et
al. 1998) type corrections which are smaller for a disk population
than for a spherical population.

The $V$ and $I$ data suggest that the slope of the Galactic
$PL$-relations may be shallower than that observed for LMC Cepheids,
either for the whole period range, or that there is a break at short
periods (near $\log P_0 \approx 0.7-0.8$).

We stress the importance of two systematic effects which influence the
distance to the LMC: the slopes of the Galactic $PL$-relations and
metallicity corrections. In order to assess the influence of these
various effects, we present 27 distance moduli (DM) to the LMC. These
are based on three different colours ($V,I,K$), three different slopes
(the slope observed for \C\ in the LMC, a shallower slope predicted
from one set of theoretical models, and a steeper slope as derived for
Galactic \C\ from the surface-brightness technique), and three
different metallicity corrections (no correction as predicted by one
set of theoretical models, one implying larger DM as predicted by
another set of theoretical models, and one implying shorter DM based
on empirical evidence). We derive DM between 18.45 $\pm$ 0.18 and
18.86 $\pm$ 0.12. The DM based on $K$ are shorter than those based on
$V$ and $I$ and range from 18.45 $\pm$ 0.18 to 18.62 $\pm$ 0.19, but
the DM in $K$ could be systematically too low by about 0.1 magnitude
because of a bias due to the fact that NIR photometry is available
only for a limited number of stars.

From the Wesenheit-index we derive a DM of 18.60 $\pm$ 0.11, assuming
the observed slope of LMC \C\ and no metallicity correction, for want
of more information.

The DM to the LMC based on the parallax data can be summarised as
follows. Based on the $PL$-relation in $V$ and $I$, and the
Wesenheit-index, the DM is 
\begin{displaymath}
18.60 \pm 0.11 \;\; (\pm 0.08 \;{\rm slope})(^{+0.08}_{-0.15} \;{\rm
metallicity}), 
\end{displaymath}
which is our current best estimate.
Based on the $PL$-relation in $K$ the DM is $\;\;\;\; 18.52 \pm 0.18$
\begin{displaymath}
 \;\;(\pm 0.03 \;{\rm slope}) (\pm 0.06 \;{\rm metallicity})
(^{+0.10}_{-0} \;{\rm sampling \;bias}). 
\end{displaymath}
The random error is mostly due to the given accuracy of the \HP\
parallaxes and the number of Cepheids in the respective samples.  The
terms between parentheses indicate the possible systematic
uncertainties due to the slope of the Galactic $PL$-relations, the
metallicity corrections, and in the $K$-band, due to the limited
number of stars. Recent work by Sandage et al. (1999) indicates that
the effect of metallicity towards shorter distances may be smaller in
$V$ and $I$ than indicated here.

From this, we point out the importance of obtaining NIR photometry for
more (closeby) \C, as for the moment NIR photometry is only available
for 27\% of the total sample. This would eliminate the possible bias
due to the limited number of stars, and would reduce the random error
estimate from 0.18 to about 0.10 mag. Furthermore, the sensitivity of
the DM to reddening, metallicity correction and slope are smallest in
the $K$-band.


\end{abstract}

\section{Introduction}

Cepheids are important standard candles in determining the
extra-galactic distance scale. The results of the \HP\ mission allow,
in principle, a calibration of the period-luminosity relation based on
the available parallaxes. Feast \& Catchpole (1997; hereafter FC) did
just that for the $M_{\rm V} - \log P$-relation based on pre-released
\HP\ data of 223 Cepheids available to them at that time. Now that the
entire catalog has become available (ESA 1997) it is timely to analyse
the full sample of Cepheids in it.

In a recent paper, Lanoix et al. (1999, hereafter L99) presented a
study similar to ours and they derived the zero points of the $M_{\rm
V} - \log P$- and $M_{\rm I} - \log P$-relations, without, however,
discussing the distance to the LMC. We will indicate where the two
studies agree and differ.

The paper is organised as follows. In Sect.~2 the sample considered in
the present paper is presented, and compared to that in FC and L99.
In Sect.~3 several aspects involved in the analysis of parallax data
are described, and the method of ``reduced parallaxes'' is outlined,
together with all necessary recipes to obtain the reddening. In
Sect.~4 the zero points of the $PL$-relations in $V,I, K$ and the
reddening-free ``Wesenheit-index'' (e.g. Tanvir 1999, and Eq.~(11)) are
presented for different selections of the sample, which are discussed
in Sect.~5. In Sect.~6 we construct and present the zero points for
volume complete samples of stars. In Sect.~7 we describe numerical
simulations that are first of all tuned to fit the observed properties
of the \C\ in the \HP\ catalog, and then are used to show that the
method of ``reduced parallaxes'' introduces a bias which is of the
order of 0.01 mag or less.  Based on these results we discuss in
Sect.~8 the distance to the LMC, and elaborate on the various
uncertainties.  \\

\section{Sample selection}

The number and some properties of the Cepheid population in the \HP\
catalog were discussed by Groenewegen (1999, hereafter G99). To
summarise: by cross-correlating the general \HP\ database, the \HP\
``resolved variable catalog'' and the electronic database of Fernie et
al. (1995; hereafter F95), a total of 280 \C\ was identified. Then, 9
Type {\sc ii} \C, 1 factual RR Lyrae variable, 1 CH-like carbon-rich
Cepheid, 10 double-mode \C, 7 Cepheids with an unreliable \HP\
solution and 4 \C\ with no or unreliable optical photometry were
excluded. Note that for RY Sco and Y Lac we use the new determinations
for the parallax and its error from Falin \& Mignard (1999).  This
leaves 248 stars, of which 32 are classified as overtone pulsators by
F95, Antonello et al. (1990) or Sachkov (1997). Luri et al. (1999)
have classified \C\ in fundamental and overtone pulsators using the
\HP\ lightcurves, but did not yet publish the results for individual
stars. For the sample, G99 calculated intensity-mean $I$ (on the
Cousins system) and $V-I$ magnitudes for 189 stars, and collected
magnitude-mean colours for additional 14 stars, and provided $JHK$
intensity-mean magnitudes on the Carter system for 69 stars.  \\

By comparison, the FC dataset consisted of 223 stars of which 3 were
discarded.  These were DP Vel for lack of photometric data, and AW Per
and AX Cir because they are in binaries and the photometry might be
affected by the companions. However, in the F95 catalog there are many
more stars which are flagged for this reason. From the sample in G99
are therefore excluded the stars flagged ``O C'' (RX Cam, AW Per, T
Mon, SS CMa, S Mus, AX Cir, W Sgr, V350 Sgr, U Aql, SU Cyg, V1334 Cyg)
and ``O: C'' (VY Car) in F95. This leaves 32 overtone pulsators and
204 fundamental mode pulsators in the sample considered here. They are
listed for completeness in Appendix A together with some adopted
parameters. \\

Recently, L99 also studied the \C\ in the \HP\ catalog. They selected
stars listed as ``DCEP'' or ``DCEPS'' from the \HP\ catalog. 
Interestingly, they state that they selected 247 stars, while in
reality there are 250 such stars (G99).  After removing 9 \C\ for
which there is no photometry listed in F95, their final sample
consists of of 238 stars (including 31 overtones). They did not
eliminate double-mode \C, \C\ where the photometry is (likely)
contaminated by a binary companion or \C\ with unreliable \HP\
solutions. Furthermore, they assumed all \C\ classified as ``DCEPS''
to be overtone pulsators, which is not the case. Their sample
therefore includes stars they consider overtones which we and FC
consider fundamental mode pulsators (e.g., SZ Cas, Y Oph, V496 Aql,
V924 Cyg, V532 Cyg), and stars they consider fundamental mode
pulsators which are overtone pulsators (e.g., V465 Mon, DK Vel, V950
Sco, see Mantegazza \& Poretti 1992). In addition, V473 Lyr is
considered by them to be a first overtone pulsator, while it probably
is a second overtone pulsator (Van Hoolst \& Waelkens 1995;
Andrievsky et al. 1998; also see below). \\

Of the 236 stars in our sample there are 198 in common with the sample
of FC. In other words, 22 stars in the FC sample would not have made
it through the selection process outlined in G99 and
here. Specifically, their sample contains 7 stars with unreliable \HP\
solutions (we use the improved parallax values for RY Sco and Y Lac
from Falin \& Mignard (1999), information which was not available to
FC), 9 binaries (in addition to AW Per and AX Cir) where the
photometry may be contaminated by the companions and 8 double-mode
\C. Another difference is that 7 more stars than in FC are flagged as
overtone pulsators following F95 and Sachkov (1997). \\

For the 198 stars in common we have compared the intensity-mean $V$
and $B-V$. FC mention that they use F95, except when the data were in
Laney \& Stobie (1993). Our photometry was at first instance taken
solely from F95, except for RW Cas (see discussion in G99). In $V$,
the photometry is identical for 165 stars. For 18 stars $V$ differs by
$>0.01$mag, for 8 by $>0.02$mag  and for 5 by $>0.03$mag. The latter
cases were inspected individually.

For RW Cas the difference in the sense ``Fernie et al $-$ FC'' is 0.101 mag. 
As discussed in G99, there may be a typographical error in F95. 

For X Pup the difference is $-0.047$. The F95 entry of $V$ =
8.460 is close to the value derived in G99 for the dataset of Moffett
\& Barnes (1984). The other dataset considered in G99 gives a value of
$V = 8.536$. The value by FC of $V = 8.507$ is intermediate. We have
kept the F95 value.

For AQ Pup the difference is 0.122 mag. G99 calculated $V,I$ for 3
data sets. The entry in F95 ($V = 8.791$) is identical to the Moffett
\& Barnes (1984) dataset. The FC value of $V = 8.669$ is close to the
2 other datasets (8.676, 8.686 mag), and one of these was the
preferred one in G99 regarding the $V,I$ photometry. For AQ Pup we use
the $V$ and $B-V$ from FC.

For RS Pup the difference is $-0.081$. The value in F95 ($V$ = 6.947)
is clearly off from both values in G99 (6.999 and 7.020 mag)  which
both are in agreement with the value of $V$ = 7.028 used by FC. For RS
Pup we use $V$ and $B-V$ from FC.

For S Nor the difference is $-0.032$. The value used by FC is
extremely close to the value derived in G99, and for S Nor we use the
$V$ and $B-V$ from FC.

After these changes in $V$ and $B-V$, there are 164 with identical
values for $B-V$, for 12 stars $B-V$ differs by $>0.014$ mag, for 9
by $>0.028$ mag, and for 5 by $>0.042$ mag (RW Cas, X Pup, U Nor, RY
Sco, RU Sct).  These cases have not been considered separately. It
merely indicates that errors in $V$ and $B-V$ (and hence reddening)
can contribute to the uncertainties in the derivation of the zero
point of the $PL$-relation, as will be discussed below.

For V1162 Aql, we discovered an error in the $(B-V)$ value listed in F95.
From the data in Berdnikov \& Turner (1995) that was used in G99 to
calculate $(V-I)$ we derive and adopt an intensity weighted mean of
$(B-V)$ = 0.879 magnitudes.

In G99, $V$ and $I$ photometry was presented for many \C\ in the \HP\
catalog, based on a literature search. When possible, intensity-mean
magnitudes were calculated based on the original published
datasets. Some magnitude-means were also presented. The $V - I$
magnitudes are taken from G99. When there are multiple entries the
first one was taken following the considerations in G99. In total
there is $I$-band data for 191 stars (or 81\% of the sample), 178 of
which are intensity-mean magnitudes. In G99 it was shown that there is
no significant difference between the intensity-mean and
magnitude-mean, and so the remaining 13 magnitude-means are used
without correction.

G99 also presented intensity-mean $JHK$ magnitudes in, or transformed
to, the Carter system. When there are multiple entries the first one
is taken, following the considerations in G99. In total there is
$JHK$-band data for 63 stars, or 27\% of the sample.

\section{Analysing parallax data}

Analysing parallax data is not a trivial exercise, and has led to some
confusion in the literature.  Part of this confusion is probably
related to the fact that parallax data suffer from many types of
biases (see Brown et al. 1997).  For example, the most conspicuous
bias, the Lutz-Kelker (LK; Lutz \& Kelker 1973) bias, although known
for a long time, could not be empirically investigated due to a lack
of data on which it could be tested.

One can visualize the LK effect by analogy with the well-known
Malmquist bias. Malmquist bias occurs on samples of objects because
due to a particular magnitude cut, objects that are brighter than the
mean will be included in a sample, while objects that are fainter than
the mean will be excluded. The net effect of this type of bias is that
the intrinsic magnitude for a certain sample will be too bright if no
corrections are applied.  To investigate the presence of such biases,
one can look at so-called Spaenhauer diagrams, where the observed
magnitude is plotted as a function of, for example, distance. In such
a way it is relatively easy to find the point where the Malmquist bias
starts to dominate (see e.g. Sandage 1994).

A similar, but opposite effect, is due to the LK bias. For a given
parallax cut (or any selection based on the parallax, or its
associated error), on average, more objects that are located far away
will scatter into the sample, than stars will scatter out of the
sample, as the sampled volume is larger for the former objects. Because 
the distances to the objects scattered into the sample are then
underestimated, this effectively results in a too faint mean
`intrinsic' magnitude of the entire sample. As with the Malmquist
bias, the LK-bias depends on the space distribution of the sample of
stars under consideration. Contrary also to the Malmquist bias is that
although the error-bars on the parallax are symmetric, those on the
derived distances and intrinsic magnitudes are a-symmetric, which
seriously affects the analysis of the data. In addition, the relative
error-bars on the parallax increase with distance.

The LK bias was shown to exist empirically by Oudmaijer et al. (1998).
Figure~1 gives an illustrative plot, showing the magnitudes of the
Cepheids under consideration, derived from their photometry and
observed parallax, as a function of parallax in the upper panel, and
as function of relative error in the lower panel ($\rho = V_0 + 5 \log
\pi +5 - \delta \log P$, for $\delta = -2.81$ -- see next Section for
details). As the absolute error in the \HP\ data is more or less
constant, the two figures are equivalent.  The inferred magnitude of
the objects in the Cepheid sample is not a random distribution around
the mean, but shows a clear trend as function of parallax. For large
parallaxes (and small relative errors) the inferred magnitudes are too
faint, and then, for larger distances, the objects become too
bright. It was only in Oudmaijer et al. (1998), that this was shown
empirically for the first time.


\begin{figure}
\centerline{\psfig{figure=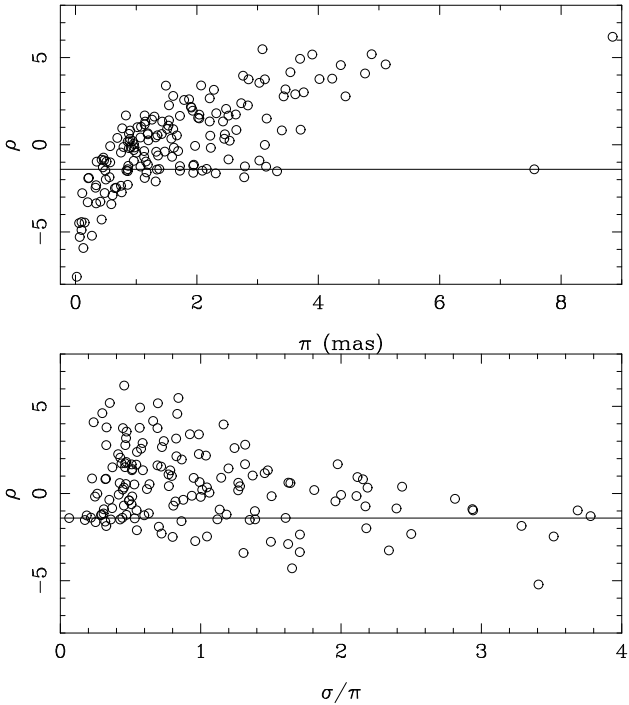,width=8.5cm,angle=-90}}
\caption[]{$\rho = V_0 + 5 \log \pi +5 - \delta \log P$ plotted
against parallax, and relative parallax error, for the stars with
positive observed \HP\ parallaxes. For comparison the derived value
for the zero point $\rho$ derived for the full sample (Table~1,
solution 10) is indicated.}
\end{figure}

Indeed, it is clear from this type of graph, which is a version of a
Spaenhauer diagram, that a selection on parallax (or relative error in
parallax) will not return the correct answer. In most cases the result
will be  biased (but see Oudmaijer et al. 1999, for a counter-example).

Lutz \& Kelker (1973) were the first to investigate this effect, and
found that the resulting bias can be substantial, increasing with
increasing relative error. For a relative error of 17.5\% in the
parallax, the bias is 0.43 magnitudes -- yet as Koen (1992) later
showed, the confidence intervals around these corrections are large.
The 17.5\% limit was taken literally by subsequent authors, and often
stars with worse determinations were deleted
or individual stars were corrected with the
values LK derived.

LK claimed that individual stars suffer from the same bias, a point
they emphasize strongly.  Actually, it is quite a surprising result
that LK concluded that individual stars are biased.  LK start their
calculations assuming that the observed parallax equals the true
parallax with a random error, a standard procedure when investigating
measurement errors and the main assumption for Monte-Carlo simulations
as performed in e.g. Sect.~7 of this paper. It seems rather
contradictory then that LK infer from their results that individual
stars are biased if their initial conditions assume otherwise.

Let us first comment on whether the conjecture that the observed
parallax is equal to the true parallax (plus a random observational
error) is correct.  It is actually quite hard to investigate this in a
`bias-free' manner, but the evidence provided by the \HP\ mission
swings the balance towards answering this question with a
`yes'. Measurements of stars beyond the detection limit of \HP, such
as those in the Magellanic Clouds (Van Leeuwen 1997, Arenou et
al. 1995) consistently give average observed parallaxes close to zero
mas with a scatter of order the observational error. This indicates
that the measurements are not biased, even though the relative errors
on the parallax are substantial -- and the results for individual
stars can be widely off the true distance, but crucially, in the mean
they appear to give the right answer within the errors rather than 
show a conspicuous bias in one direction. In addition, analyses of
Open Clusters, for which photometric distances are known, do not show
a deviation from the mean distance for increasing distances and thus
lower quality data (Arenou et al. 1995, Robichon et al. 1999).  Based
on this, it appears that individual stellar determinations are most
likely not biased.

There is one exception, mentioned briefly by Koen \& Laney
(1998). These authors argue that if a single star is the result of a
selection on parallax, its measurement is biased. Although this sounds
counter-intuitive, it is a formal implication from the fact that a
selection on parallax gives rise to a bias.


A final comment concerns the use of parallaxes of a sample of which
one does not know that the objects should have the same intrinsic
magnitude, while their distances are not known.
The trend observed in Fig.~1 that, for a
given sample, large parallaxes return in principle too faint intrinsic
magnitudes, while small parallaxes return too bright magnitudes, can
have serious implications in the interpretation of the data.

\subsection{How to deal with the Lutz-Kelker bias?}

The remaining question is how one should deal with the effects of the
bias. Should one, as often the case with Malmquist bias, introduce a
(sample-dependent) correction factor, or are there ways to circumvent
the problem?

In a paper dealing with this problem, Turon Lacarrieu \& Cr\'ez\'e
(1977) discuss two different methods: \footnote{Note that Turon
Lacarrieu \& Cr\'ez\'e (1977) repeat the phrase by LK that `individual
stars' suffer from the bias, without specifying whether this would
mean {\it all} stars (which contradicts their own assumption that the
observed parallax is equal to the true parallax with an associated
error-bar), or individual stars that are the result of a selection on
parallax. }

Accepting that a sample selected on parallax is inevitably biased,
they first considered using the better parallaxes, and investigate the
resulting mean magnitudes, and provide corrections along similar lines
as LK (also see Smith 1987a+b+c, Koen 1992).

Secondly they considered a full sample, and, avoiding transformations
from parallaxes to distances and magnitudes, they derive the mean
parallax first and use the resulting mean parallax to derive the mean
magnitude of the sample. Since negative parallaxes can not be
incorporated into a mean magnitude, one in principle has to discard
these data, in effect selecting on parallax and thus biasing the
sample. Therefore, Turon Lacarrieu \& Cr\'ez\'e (1977) introduced the
so-called ``reduced parallax'' (10$^{0.2M_{\rm V}} \sim \pi$), which
can take into account negative parallaxes, and hence the (weighted)
mean of the reduced parallax can be converted into a mean magnitude.
This method, as will be discussed in Sect.~3.2, indeed appears to be
``bias-free'', mainly because a weighting scheme puts less weight on
the larger deviations around the mean from the lower quality data,
whilst not being a formal selection on parallax.

The Cepheids in \HP\ were investigated previously by FC who used the
second, reduced parallax method, and Oudmaijer et al. (1998), who used
a scheme based on the first method. Their results were equal within
the error-bars. Although FC used the reduced parallax method, which
formally does not suffer from LK bias (Koen \& Laney 1998), they still
suggested in a footnote that a LK correction of 0.02 mag should be
applied on their final result. This is not necessary; the result only
has to be corrected for Malmquist bias, as FC also pointed out, but
did not actually apply, as they estimated it would essentially
counteract their proposed LK correction of 0.02 mag.

These, and other issues will be investigated in the remainder of this
paper. First, we will outline the method for the reduced parallaxes again.

\subsection{The ``reduced parallax'' method}

The method of ``reduced parallax'' (discussed by Turon Lacarrieu \&
Cr\'ez\'e 1977) was the one used by FC in analysing Cepheid data. \\

\noindent
Consider a Period-Luminosity relation of the form:
\begin{equation}
 M_{\rm V} = \delta \, \log P + \rho,
\end{equation}
where $P$ is the fundamental period in days. If $\langle V\rangle$ is
the intensity-mean visual magnitude and $\langle V_0 \rangle$ its
reddening corrected value, then one can write:
\begin{equation}
10^{0.2\rho} = \pi \times 0.01\,\;10^{0.2(\langle V_0 \rangle - \delta
\;\log \;P)} \equiv \pi \times {\rm RHS},
\end{equation}
which defines the expression {\sc rhs} and where $\pi$ is the parallax
in milli-arcseconds. This method has the advantage that negative
parallaxes can be used in the analyses as well. A weighted-mean, with
error, of the quantity 10$^{0.2 \rho}$ is calculated, with the weight
(weight = $\frac{1}{{\sigma}^2}$) for the individual stars derived
from:
\begin{equation}
{\sigma}^2 = \left( {\sigma}_{\pi} \times {\rm RHS} \right)^2 + 
\left(0.2\,\ln(10) \,\; \pi\; {\sigma}_{\rm H}  \times {\rm RHS} \right)^2,
\end{equation}
with ${\sigma}_{\pi}$ the standard error in the parallax. This follows
from the propagation-of-errors in Eq.~(2). For the error (denoted
${\sigma}_{\rm H}$) in $(\langle V_0 \rangle-\delta \; \log \;P)$ we
follow FC's ``solution B'' and adopt ${\sigma}_{\rm H} = 0.1$
throughout this paper. Recently, L99 considered alternative weighting
schemes, but concluded that the one used by FC and the present paper
gives the most reliable zero point and the lowest dispersion. \\

\noindent
The reddening is derived as follows. The intrinsic colours follow from
the relation in  Laney \& Stobie (1994):
\begin{equation}
  \langle B \rangle_0 - \langle V \rangle_0 \;= 0.416 \, \log P + 0.314,
\end{equation}
which has a dispersion of 0.091 mag. The visual extinction
($A_{\rm V} = R_{\rm V} \times E(B-V)$) is calculated using (Laney
\& Stobie 1993):
\begin{equation}
  R_{\rm V} = 3.07 + 0.28 \; (B-V)_0 + 0.04\; E(B-V). 
\end{equation}
No dispersion is given for this relation, only an error of 0.03 in the
zero point. We will assume that the dispersion in Eq.~(5) is 
slightly larger than this, namely 0.05 mag.

\noindent
For overtone pulsators, the fundamental period has to be estimated
from the observed period. This was done, following FC, using:
\begin{equation}
  P_1/P_0 = 0.716 - 0.027\; \log P_1,
\end{equation}
with a dispersion we estimate from the original data (Alcock et
al. 1995) to be of order 0.002.

\noindent
For pulsators in the second overtone, the fundamental period is
calculated, following FC, using:
\begin{equation}
  P_2/P_0 = 0.55.
\end{equation}
This completes the description of the method used by FC.  \\

\noindent
Alternative methods which are described now, follow this description
closely but are based on $\langle V \rangle$ and $\langle I \rangle$,
respectively $\langle J \rangle$ and $\langle K \rangle$ photometry
instead of $\langle B \rangle$ and $\langle V \rangle$. The rationale
being that the extinction in $I$ and $K$ is less than in $V$, and that
the scatter in the $M_{\rm I}-P$- and $M_{\rm K}-P$-relations is less
than in the $M_{\rm V}-P$-relation (Tanvir 1999, Laney \& Stobie
1994).  \\

\noindent
First consider the case based on $V$ and $I$. The intrinsic colour is
derived from (Caldwell \& Coulson 1986):
\begin{equation}
  \langle V \rangle_0 - \langle I \rangle_0 \;= 0.292 \, \log P + 0.443
\end{equation}
which has a dispersion of 0.064 mag.  This relation was derived for
magnitude-mean $(V-I)_0$ but we will assume it to hold for
intensity-mean magnitudes as well. In a recent paper, Feast (1999,
Appendix D), presents a correction formula that implies that the
difference intensity-mean minus magnitude-mean $(V-I)$ colour 
is of order $-0.017$ mag for a typical $V$-band amplitude of 0.7 mag. \\

\noindent
There seems not to exist a relation similar to Eq.~(5), where $R(I)
\equiv A_{\rm I}/E(V-I)$ is related to $(\langle V \rangle-\langle I
\rangle)_0$ and/or $E(V-I)$. We have derived such a relation from the
available $BVI$ data.

\noindent
For each star with $BVI$ photometry, $(\langle B \rangle-\langle V
\rangle)_0$ and $(\langle V \rangle-\langle I \rangle)_0$ can be
calculated from Eqs.~(4) and (8). Then, $A_{\rm I}$ is calculated
using Gieren et al. (1998):
\begin{equation}
  R_{\rm I} = 1.82 + 0.205 \; (B-V)_0 + 0.022\; E(B-V),
\end{equation}
and $A_{\rm I} = R_{\rm I} \times E(B-V)$.  In Fig.~2 $R(I)$ is
plotted versus $(V-I)_0$. A least-square fit to 183 data points gives:
\begin{equation}
  R(I) = 1.422
\end{equation}
with no significant dependence on $(V-I)_0$ and with a standard error
of 0.19. For ${\sigma}_{\rm H}$ (in this case the error in $(\langle
I_0 \rangle-{\delta}_{\rm I} \; \log \;P$)) we adopt a value of 0.15
mag (Tanvir 1999). \\


\noindent
A variation on this method that treats the problem of reddening in a
different way, is to use the reddening-free so-called
``Wesenheit-index'' (see for example Tanvir 1999), that uses the {\em
observed} colours but is essentially reddening-free when defined as:
\begin{equation}
  W = V - 2.42\; (V-I).
\end{equation}
For ${\sigma}_{\rm H}$ (in this case the error in $(\langle W
\rangle-{\delta}_{\rm W} \; \log \;P$)) we adopt a value of 0.11
mag (Tanvir 1999). \\

\noindent
Now consider the case based on $J$ and $K$ colours. The intrinsic color is
derived from Laney \& Stobie (1994):
\begin{equation}
  \langle J \rangle_0 - \langle K \rangle_0 \;= 0.149 \, \log P + 0.310
\end{equation}
which has a dispersion of 0.044 mag. Again, there seems not to exist a
relation similar to Eq.~(5), where $R(K) \equiv A_{\rm K}/E(J-K)$ is
related to $(\langle J \rangle-\langle K \rangle)_0$ and/or
$E(J-K)$. We have derived such a relation from the available $BVJK$
data.

\noindent
For each star with $BVJK$ photometry, $(\langle B \rangle-\langle V
\rangle)_0$ and $(\langle J \rangle-\langle K \rangle)_0$ can be
calculated from Eqs.~(4) and (12). Then, $A_{\rm K}$ is calculated
using Laney \& Stobie (1993):
\begin{equation}
  A_{\rm K} = 0.279 \; E(B-V).
\end{equation}
In Fig.~3 $R(K)$ is plotted versus $(J-K)_0$. A least-square fit to 55
data points gives:
\begin{equation}
  R(K) = 1.035 - 1.063 \; (J-K)_0 
\end{equation}
with a standard error of 0.091.  For ${\sigma}_{\rm H}$ (in this case
the error in $(\langle K_0 \rangle-{\delta}_{\rm K} \; \log \;P$)) we
adopt a value of 0.12 mag (Gieren et al. 1998).


\begin{figure}
\centerline{\psfig{figure=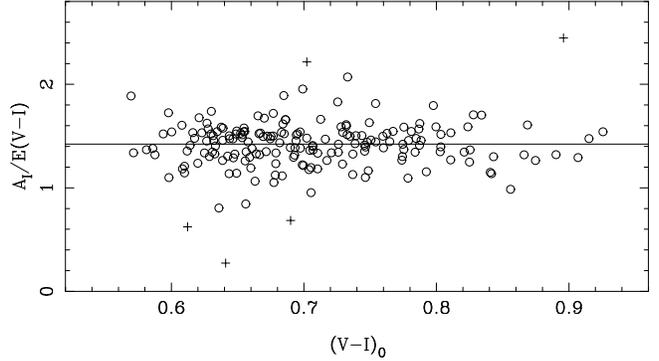,width=8.8cm,angle=-90}}
\caption[]{The relation between $R(I)$ and $(\langle V \rangle-\langle
I \rangle)_0$. The fit reported in Eq.~(10), is based on 183
stars. Eight outliers (the crosses) were not considered in the
fit. Three outliers are outside the plot range.  There is no
dependence on $(V-I)$.  }
\end{figure}

\begin{figure}
\centerline{\psfig{figure=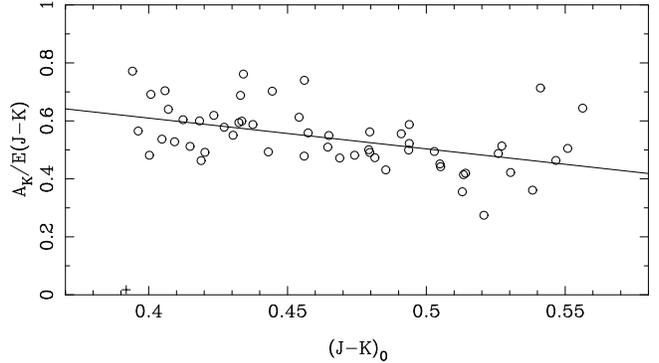,width=8.8cm,angle=-90}}
\caption[]{The relation between $R(K)$ and $(\langle J \rangle-\langle
K \rangle)_0$. The fit reported in Eq.~(14), is based on 55
stars. Eight outliers (the crosses) were not considered in the
fit. Seven outliers are outside the plot range.  There is a dependence
on $(J-K)$.  }
\end{figure}

\section{Zero points of the $PL$-relations}

\subsection{Zero point of the $M_{\rm V}-\log P$-relation}

In Tables~1-4 we present the results from the ``reduced parallax''
method presented in the previous section for different samples of
stars, and based on different colors ($BV$, $VI$ or $JK$). To test the
implementation of the method we have calculated the zero point for
some of the solutions considered in FC, adopting a slope $\delta =
-2.81$ and working with $BV$ colours, as they did. The data for the
sample used by FC come from Feast \& Whitelock (1997; their
Table~1). These are our solutions 1-7 in Table~1, and the value for
the zero point and total weight are in perfect agreement to within the
listed number of decimal places in FC. The error in the zero point
determination is slightly different, ours being larger by a few
1/100-th of a magnitude.  Using the FC sample, we also calculated the
solution for the overtones only, and the overtones excluding Polaris
(solutions 8-9). \\

For the present sample and using a slope $\delta = -2.81$ solutions
10-12 give the value for the zero point for the whole sample, and for
fundamental mode and overtone pulsators separately. In this case, V473
Lyr was considered to be pulsating in the second overtone. This
interesting object is thought to be the only Galactic Cepheid
pulsating in the second overtone (Van Hoolst \& Waelkens 1995 and
Andrievsky et al. 1998). The values of $\rho$ assuming that V473 Lyr
is a fundamental mode, first overtone or second overtone pulsator,
respectively, are $-2.68$, $-2.07$ and $-1.61$, all with an error of
0.70. With such a large error only fundamental mode pulsation can be
excluded from the \HP\ parallax alone. The zero point assuming second
overtone pulsation is consistent with the zero point obtained for the
whole sample, and we will hence assume in our zero point
determinations that V473 Lyr is indeed a second overtone pulsator. \\

The zero point for the entire sample of $-1.41 \pm 0.10$ compares to
the value of FC of $-1.43 \pm 0.10$, and the value of L99 of $-1.44
\pm 0.05$ (they used a slightly different slope of $-2.77$). These
values are all very similar, and the differences are mainly due to the
different samples of \C, and to a lesser extent to the slight
differences in the adopted photometry. We believe that the present
sample is the most carefully selected sample of the three with respect
to completeness, the flagging of overtone pulsators, and the removal
of \C\ that may influence the analysis for various reasons
(double-mode \C, unreliable \HP\ solutions, possible contaminated
photometry due to binary companions), as explained in the
introduction.

\subsubsection{$P_0$ versus $P_1$}

A first comment is on the different solution from the fundamental mode
and overtone pulsators. In FC the difference between the solution
using the fundamental pulsators or the full sample was only 0.02
mag. In our case it is about 0.08. FC did not present the solution for
the overtones only, but we have calculated it from their data
(solution 8). The difference using only the overtones or the
fundamental pulsators is 0.05 mag in their case. In our case it is
0.15 mag, which is a difference at the 1$\sigma$ level. This indicates
how important it is to carefully flag the overtone pulsators.

\subsubsection{Selecting on visual magnitude}

L99 derive a zero point of $-1.44$ (after correcting for a bias of
$-0.01$ mag) with a very small error of 0.05, using a selection on $V
\le 5.5$. We confirm (solutions 14 and 15) that the derived zero point
is not significantly different from that for the whole sample, but we
do not confirm such a small error. In fact, Pont (1999) argues that
the error of 0.10 in FC is even underestimated based on his numerical
simulations. Our simulations confirm this (Sect.~7) and so it is not
clear how L99 arrived at such a small error.

\subsubsection{Selecting on parallax}

For solutions 16-18 only positive parallaxes have been selected to
highlight the effect of LK-bias. The zero points are fainter, as
expected. Interestingly enough, the zero point for the overtone
pulsators is hardly changed. Only 5 overtones (15\%) have negative
observed parallaxes and those carry very little weight, while for the
fundamental mode pulsators a selection on positive parallaxes reduces
the number by 32\%.

\subsubsection{Selecting on weight}

FCs final choice for the zero point relied heavily on a subsample of
26 stars, selected on weight. The question arises if this introduces
some bias.  For solutions 19-24 we have made different selections on
weight, in particular solutions 19-22 have been devised such that each
bin selected on individual weight has about the same total weight, so
that the errors on the zero point are comparable. The differences are
at the 1$\sigma$ level. From numerical simulations performed in
Sect.~7 (see Table~6) we confirm that there are no indications at the
present level of accuracy that a selection on weight is an (indirect)
selection on parallax, and so a sample selected on weight appears not
to be subject to LK-bias. The error on the zero point determination is
larger when selecting on (individual) weight, simply because of the
smaller value of the total weight (see Table~6).  This is different
from the result in FC, who quote a {\it smaller} error on the zero
point for the sample selected on weight compared to their full sample
(cf. their solution 6 and 1).


\subsubsection{Selecting on period}

Solutions 25-36 represent cases where the sample was split up in bins
in $\log P$ carrying approximately equal weight, for the whole sample
(solutions 25-28), and the fundamental pulsators only (solutions
29-32, and 33-36 for a slope of $-2.22$). In both cases, the star with
the highest individual weight was excluded in defining the bins. There
is a significant dependence of the zero point on $\log P$. This is
particularly clear in the case of the fundamental mode pulsators where
the zero point for stars with $\log P \ge 0.85$ differs at the
3$\sigma$ level from the shortest period bin.

To expand on this matter further, we show in Fig.~4 how 10$^{0.2
\rho}$ depends on $\log P$ for the full sample (top panel), and for
the 47 stars with an individual weight $>5$. Shown are the
weighted-mean values of 10$^{0.2 \rho}$ (solid lines), and weighted 
least-square fits to the data of the form 10$^{0.2 \rho} = \alpha \,
\log P + \beta $ (dashed line). The ($\alpha, \beta$) found are
(0.11 $\pm$ 0.12, 0.44 $ \pm$ 0.10) for the full sample, and (0.094
$\pm$ 0.115, 0.43 $ \pm$ 0.09) for the 47 stars. The analysis was
repeated for the fundamental mode pulsators only, giving samples of
204 and 35 stars, respectively.  The ($\alpha, \beta$) found are (0.17
$\pm$ 0.15, 0.35 $ \pm$ 0.14) for the full sample, and (0.17 $\pm$
0.15, 0.34 $ \pm$ 0.14) for the stars with weight $>5$. The slopes
derived are significant at the 1$\sigma$ level only.  \\

These results depend on the adopted slope in the $PL$-relation. For a
slope $\delta = -3.1$ we find that the effect of the dependence of the
zero point on the binning in $\log P$ is enhanced, and that the slope
in the 10$^{0.2 \rho}$ versus $\log P$ relation becomes significant
at the 2$\sigma$ level. We also calculated the results for a slope of
$\delta = -2.22$ (Bono et al. 1999).  The results are listed as
solutions 33-36 for the fundamental pulsators. The dependence on
period is not significant in this case, although the shortest bin
remains the brightest. Least-square fits give the following results
for ($\alpha, \beta$): all stars ($-0.039 \,\pm$ 0.090, 0.45 $\pm$
0.08), 56 stars with weight $>5$ ($-0.050 \,\pm$ 0.099, 0.45 $\pm$
0.08), all fundamental pulsators (0.00 $\pm$ 0.11, 0.39 $\pm$ 0.11),
42 fundamental pulsators with weight $>5$ (0.00 $\pm$ 0.13, 0.38 $\pm$
0.12).  The slope derived is no longer significant. From this analysis
one may conclude that there is weak evidence for the fact that the
slope in the $M_{\rm V}-\log P$-relation is shallower than in the LMC,
or, alternatively that there is a change of slope at the short period
end. A similar effect is found for the $M_{\rm I}-\log P$-relation
(see next section). The implications are discussed further in
Sect.~8. \\

\begin{table*}
\caption[]{Values for the zero point from $BV$ photometry.}


\begin{tabular}{rrcrl} \hline
Solution& N  &   Zero point  & Total  & Remarks \\
        &    &    in $V$     & Weight &         \\ \hline
1 &220 & -1.403 $\pm$ 0.104 & 1598.1 & FC solution 1 (whole sample) \\
2 &219 & -1.399 $\pm$ 0.130 & 1002.2 & FC solution 2 (whole sample minus Polaris) \\
3 &210 & -1.427 $\pm$ 0.144 &  844.9 & FC solution 3 (whole sample minus overtones) \\
4 & 25 & -1.442 $\pm$ 0.154 &  751.0 & FC solution 4 (high weight minus Polaris) \\ 
5 & 20 & -1.499 $\pm$ 0.177 &  595.4 & FC solution 5 (high weight minus overtones) \\
6 & 26 & -1.428 $\pm$ 0.144 & 1346.9 & FC solution 6 (high weight plus  Polaris) \\
7 &  1 & -1.410 $\pm$ 0.170 &  595.9 & FC solution 10 (Polaris) \\
  &    &                    &        & \\
8 & 10 & -1.377 $\pm$ 0.149 &  753.2 & FC sample, only overtones \\
9 &  9 & -1.254 $\pm$ 0.308 &  157.3 & FC sample, overtones minus Polaris \\
  &    &                    &        & \\
10 & 236  & -1.411 $\pm$ 0.100 &  1718.6 & All stars \\
11 & 204  & -1.492 $\pm$ 0.150 &   824.5 & All fundamental modes \\
12 &  32  & -1.339 $\pm$ 0.135 &   894.1 & All overtones \\
13 &  31  & -1.201 $\pm$ 0.219 &   297.5 & All overtone minus Polaris \\
14 &  10  & -1.417 $\pm$ 0.128 &  1067.4 & $V_{\rm obs} < 5.5$ \\
15 &  19  & -1.461 $\pm$ 0.122 &  1225.0 & $V_0 < 5.5$ \\
16 & 165  & -1.248 $\pm$ 0.095 &  1656.7 & $\pi >0$, all stars \\
17 & 138  & -1.154 $\pm$ 0.134 &   764.2 & $\pi >0$, fundamental modes\\
18 &  27  & -1.332 $\pm$ 0.134 &   892.6 & $\pi >0$, overtones \\
19 & 208  & -1.219 $\pm$ 0.229 &   275.4 & weight $< 11$ \\
20 &  17  & -1.541 $\pm$ 0.267 &   272.8 & $11 \le $ weight $< 29$ \\
21 &   7  & -1.494 $\pm$ 0.252 &   292.5 & $29 \le $ weight $< 70$ \\
22 &   3  & -1.401 $\pm$ 0.247 &   281.3 & $70 \le $ weight $< 500$ \\
23 & 235  & -1.411 $\pm$ 0.124 &  1122.1 & weight $< 500$ \\
24 &  27  & -1.478 $\pm$ 0.147 &   846.7 & $11 \le $weight$ < 500$ \\
25 &  50  & -1.680 $\pm$ 0.274 &   296.3 & $\log P <$ 0.65 \\
26 &  54  & -1.251 $\pm$ 0.226 &   292.9 & $0.65 \le \log P < 0.79$, no Polaris\\
27 &  62  & -1.572 $\pm$ 0.272 &   271.6 & $0.79 \le \log P < 0.98$\\
28 &  69  & -1.165 $\pm$ 0.223 &   261.3 & $0.98 \le \log P$\\
29 &  66  & -2.016 $\pm$ 0.412 &   177.7 & $\log P < 0.73$, no $\delta$ Cep\\
30 &  40  & -1.607 $\pm$ 0.376 &   146.7 & $0.73 \le \log P < 0.85$\\
31 &  35  & -1.205 $\pm$ 0.323 &   136.9 & $0.85 \le \log P < 0.99$\\
32 &  62  & -1.254 $\pm$ 0.250 &   239.2 & $0.99 \le \log P$\\
33 &  66  & -2.367 $\pm$ 0.410 &   247.6 & $\delta = -2.22$, $\log P < 0.73$, no $\delta$ Cep\\
34 &  40  & -2.085 $\pm$ 0.376 &   227.7 & $\delta = -2.22$, $0.73 \le \log P < 0.85$\\
35 &  35  & -1.730 $\pm$ 0.323 &   222.7 & $\delta = -2.22$, $0.85 \le \log P < 0.99$\\
36 &  62  & -1.994 $\pm$ 0.252 &   446.8 & $\delta = -2.22$, $0.99 \le
        \log P$\\
37 & 204  & -2.019 $\pm$ 0.149 &  1349.2 & $\delta = -2.22$, all fundamental modes \\
38 & 236  & -1.885 $\pm$ 0.100 &  2662.2 & $\delta = -2.22$, all stars \\
39 &  47  & -1.495 $\pm$ 0.110 &  1555.4 & ${\pi}_{\rm phot} >1$ mas,
        all stars, ZP=-1.411 \\
40 &  45  & -1.481 $\pm$ 0.109 &  1549.7 & ${\pi}_{\rm phot} >1$ mas,
        all stars, ZP=-1.485 \\
41 &  35  & -1.643 $\pm$ 0.177 &   684.1 & ${\pi}_{\rm phot} >1$ mas,
        fundamental modes, ZP=-1.411 \\
42 &  27  & -1.615 $\pm$ 0.181 &   638.1 & ${\pi}_{\rm phot} >1$ mas,
        fundamental modes, ZP=-1.615 \\
43 &  12  & -1.386 $\pm$ 0.139 &   871.3 & ${\pi}_{\rm phot} >1$ mas,
        overtones, ZP=-1.411 \\
44 &  13  & -1.386 $\pm$ 0.139 &   875.9 & ${\pi}_{\rm phot} >1$ mas,
        overtones, ZP=-1.386 \\
45 &  12  & -1.434 $\pm$ 0.123 &  1160.4 & ${\pi}_{\rm phot} >1.8$ mas,
        all stars \\
46 &   8  & -1.388 $\pm$ 0.199 &   429.1 & ${\pi}_{\rm phot} >1.8$ mas,
     fundamental modes \\
47 &   3  & -1.444 $\pm$ 0.159 &   708.1 & ${\pi}_{\rm phot} >1.8$ mas,
     overtones \\
48 & 236  & -1.406 $\pm$ 0.100 &  1710.7 & as (10), $V$ larger by 0.005\\
49 & 236  & -1.434 $\pm$ 0.100 &  1755.1 & as (10), $B-V$ larger by 0.007\\
50 & 236  & -1.382 $\pm$ 0.100 &  1673.2 & as (10), $(B-V)_0$ larger by 0.009\\
51 & 236  & -1.417 $\pm$ 0.100 &  1727.7 & as (10), $R_{\rm V}$ larger by 0.05\\
52 & 107  & -1.155 $\pm$ 0.181 &   415.3 & All stars with $\log P \ge 0.846$ \\
53 & 107  & -1.816 $\pm$ 0.182 &   754.5 & $\delta = -2.22$, All
        stars with $\log P \ge 0.846$ \\
54 &  41  & -1.954 $\pm$ 0.114 &  2213.4 & ${\pi}_{\rm phot} >1$ mas,
 $\delta = -2.22$, $\log P > 0.50$ \\
55 &  42  & -1.299 $\pm$ 0.115 &  1187.4 & ${\pi}_{\rm phot} >1$ mas,
 $\delta = -3.04$, $\log P > 0.50$ \\

\hline
\end{tabular}
\end{table*}


\begin{figure}
\centerline{\psfig{figure=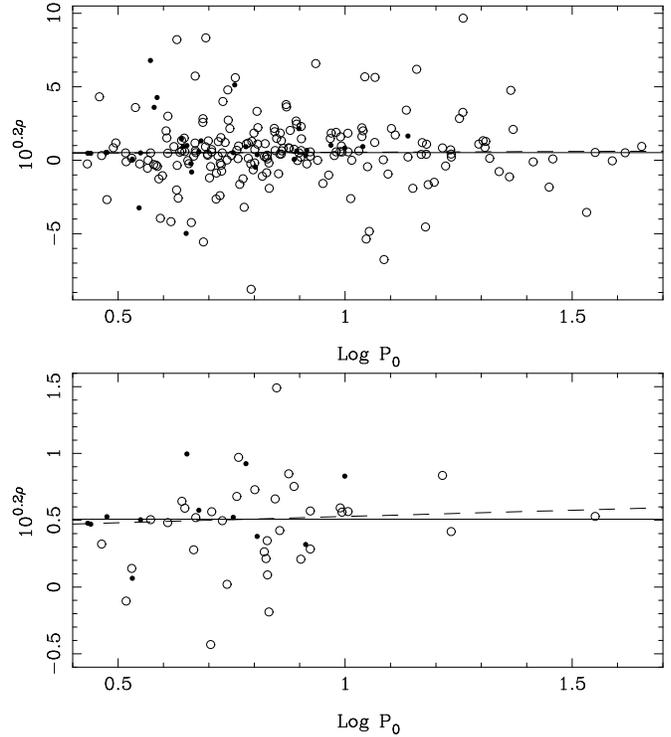,width=8.8cm}}
\caption[]{10$^{0.2 \rho}$ versus $\log P$ based on $BV$ colours for
a slope of $-2.81$. In the top panel all 236 stars are plotted. The
overtones are marked by the filled symbols. In the bottom panel only
the 47 stars with an individual weight $>5$ are shown. The solid lines
represent the weighted mean of 10$^{0.2 \rho}$ for the respective
samples. The dashed lines represent the weighted least-squares fit to
the data for the respective samples (see text).
This provides weak evidence that the slope of $-2.81$ is too steep.
}
\end{figure}

\subsection{Zero point of the $M_{\rm I}-\log P$-relation}

In Table~2 the zero points of the $M_{\rm I}- \log P$-relation are
listed based on $V,I$ photometry. Unless otherwise noted, a slope of
$-3.05$ is used. This is the slope adopted in L99, and is based on
work of Tanvir (1999) and Gieren et al. (1998). In every case we have
calculated both the zero points for the $M_{\rm I}- \log P$- and
$M_{\rm V}- \log P$-relation for the respective samples. For the whole
sample we find $\rho = -1.89 \pm 0.11$. This is 0.08 mag brighter
than in L99, but within their and our quoted errors. L99 used the
magnitude-mean magnitudes in Caldwell \& Coulson (1987), and then
applied a correction of $-0.03$ mag to convert these to
intensity-mean magnitudes. We use for most stars the intensity-mean
magnitudes as calculated in G99 from the original data. \\

As for the solutions based on $BV$ photometry we find that the zero
point using only the fundamental modes is brighter than using only the
overtone pulsators, but the difference is now less than 1$\sigma$.
Polaris again is the star with the highest individual weight.
Solutions 6-8 illustrate the effect of LK-bias when selecting stars
with positive parallaxes.

\subsubsection{Selecting on period}

As before, we have split up the sample according to period in bins of
approximate equal total weight, for all stars (solutions 9-12,
excluding Polaris), and for the fundamental pulsators (solutions
13-16, excluding $\delta$ Cep, the fundamental pulsator with the
highest individual weight).  Again we see that the zero point depends
on the period bin chosen, but the effect is not as systematic as in
the $V$-band. In Fig.~5 10$^{0.2 \rho}$ is plotted against $\log
P$. We have made least-square fits as before and find for the whole
sample ($\alpha, \beta$) = (0.11 $\pm$ 0.11, 0.331 $\pm$ 0.091), for
the 48 stars with an individual weight $>5$ ($\alpha, \beta$) = (0.101
$\pm$ 0.098, 0.332 $\pm$ 0.079), for all 163 fundamental mode
pulsators ($\alpha, \beta$) = (0.17 $\pm$ 0.14, 0.26 $\pm$ 0.13), and
for the 35 fundamental mode pulsators with a weight $>5$ ($\alpha,
\beta$) = (0.17 $\pm$ 0.12, 0.25 $\pm$ 0.11). The slope is again
significant at at most the 1$\sigma$ level. \\

Solutions 17-21 give the division in period bins for a slope of the
$PL$-relation of $-2.35$ for the fundamental mode pulsators (the slope
in the $PL$-relation in $V$ is $-2.22$). After taking into account the
difference in zero points in $V$, the dependence of the zero point in
$I$ on period bin is not significant.  The least-square fits give the
following results for ($\alpha, \beta$): the whole sample ($-0.024
 \pm$ 0.076, 0.343 $\pm$ 0.065), for the 62 stars with an individual
weight $>5$ ($-0.013$ $\pm$ 0.077, 0.317 $\pm$ 0.069), for all 163
fundamental mode pulsators (0.012 $\pm$ 0.093, 0.294 $\pm$ 0.089), and
for the 50 fundamental mode pulsators with a weight $>5$ (0.032 $\pm$
0.093, 0.264 $\pm$ 0.090). As we found previously, a shallower slope
gives rise to a smaller or no dependence of the zero point on period. \\

\begin{table*}
\caption[]{Values for the zero point from $VI$ photometry.}
\begin{tabular}{rrcrcl} \hline
Solution& N  &   Zero point  & Total   &  Zero point  & Remarks \\
        &    &   in $I$      & Weight  &  in $V$      &         \\ \hline
1 & 191 & -1.892 $\pm$ 0.111 &  2203.4 & -1.42 $\pm$ 0.10 & All stars \\
2 & 163 & -1.952 $\pm$ 0.159 &  1131.9 & -1.49 $\pm$ 0.16 & All fundamental modes \\
3 &  28 & -1.830 $\pm$ 0.154 &  1071.5 & -1.36 $\pm$ 0.14 & All overtones \\
4 &   1 & -1.859 $\pm$ 0.204 &   629.3 & -1.41 $\pm$ 0.17 & Polaris \\
5 &  10 & -1.889 $\pm$ 0.142 &  1328.4 & -1.42 $\pm$ 0.13 & $V_{\rm obs} < 5.5$ \\
6 & 139 & -1.745 $\pm$ 0.105 &  2137.9 & -1.29 $\pm$ 0.10 & $\pi >0$, all stars \\
7 & 114 & -1.671 $\pm$ 0.143 &  1068.9 & -1.22 $\pm$ 0.14 & $\pi >0$, fundamental modes\\
8 &  25 & -1.821 $\pm$ 0.154 &  1069.0 & -1.35 $\pm$ 0.14 & $\pi >0$, overtones\\
9 &  35 & -1.974 $\pm$ 0.270 &   398.0 & -1.51 $\pm$ 0.27 & $\log P < 0.66$ \\
10&  36 & -2.039 $\pm$ 0.284 &   383.4 & -1.50 $\pm$ 0.28 & $0.66 \le
        \log P < 0.78$, no Polaris \\
11&  44 & -2.131 $\pm$ 0.286 &   409.5 & -1.61 $\pm$ 0.29 & $0.78 \le \log P < 0.92$  \\
12&  75 & -1.514 $\pm$ 0.223 &   383.2 & -1.14 $\pm$ 0.22 & $\log P \ge 0.92$ \\
13&  38 & -2.148 $\pm$ 0.397 &   216.6 & -1.72 $\pm$ 0.40 & $\log P < 0.72$ \\
14&  35 & -2.408 $\pm$ 0.451 &   213.4 & -1.99 $\pm$ 0.46 & $0.72 \le
        \log P < 0.85$, no $\delta$ Cep \\
15&  29 & -1.731 $\pm$ 0.351 &   188.7 & -1.21 $\pm$ 0.34 & $0.85 \le \log P < 0.99$ \\
16&  60 & -1.629 $\pm$ 0.258 &   318.7 & -1.26 $\pm$ 0.25 & $\log P \ge 0.99$ \\
17&  38 & -2.581 $\pm$ 0.396 &   323.7 & -2.09 $\pm$ 0.40 & $\delta = -2.35$, $\log P < 0.72$ \\
18&  35 & -2.976 $\pm$ 0.451 &   358.9 & -2.47 $\pm$ 0.46 & $\delta = -2.35$, $0.72 \le
        \log P < 0.85$, no $\delta$ Cep \\
19&  29 & -2.349 $\pm$ 0.349 &   335.1 & -1.73 $\pm$ 0.34 & $\delta = -2.35$, $0.85 \le \log P < 0.99$ \\
20&  60 & -2.476 $\pm$ 0.258 &   691.6 & -2.00 $\pm$ 0.25 & $\delta = -2.35$, $\log P \ge 0.99$ \\
21& 163 & -2.571 $\pm$ 0.158 &  2020.5 & -2.03 $\pm$ 0.16 & $\delta =
        -2.35$, all fundamental modes \\
22& 191 & -2.454 $\pm$ 0.110 &  3709.3 & -1.90 $\pm$ 0.10 & $\delta = -2.35$, all stars\\
23&  12 & -1.924 $\pm$ 0.139 &  1442.5 & -1.43 $\pm$ 0.13 & ${\pi}_{\rm phot} >1.8$ mas, all stars \\
24& 191 & -1.887 $\pm$ 0.111 &  2193.3 & -1.42 $\pm$ 0.10 & as (1), $I$ larger by 0.005\\
25& 191 & -1.899 $\pm$ 0.111 &  2217.6 & -1.42 $\pm$ 0.10 & as (1), $V-I$ larger by 0.007\\
26& 191 & -1.883 $\pm$ 0.111 &  2186.1 & -1.42 $\pm$ 0.10 & as (1), $(V-I)_0$ larger by 0.006\\
27& 191 & -1.922 $\pm$ 0.111 &  2262.5 & -1.42 $\pm$ 0.10 & as (1), $R(I)$ larger by 0.19\\
28&  10 & -2.469 $\pm$ 0.143 &  2232.3 & -1.92 $\pm$ 0.13 & ${\pi}_{\rm phot} >1.8$ mas, $\delta = -2.35$, $\log P > 0.50$ \\
29&  11 & -1.689 $\pm$ 0.144 &  1082.8 & -1.23 $\pm$ 0.13 & ${\pi}_{\rm phot} >1.8$ mas, $\delta = -3.33$, $\log P > 0.50$ \\
\hline
\end{tabular}
\end{table*}

\begin{figure}
\centerline{\psfig{figure=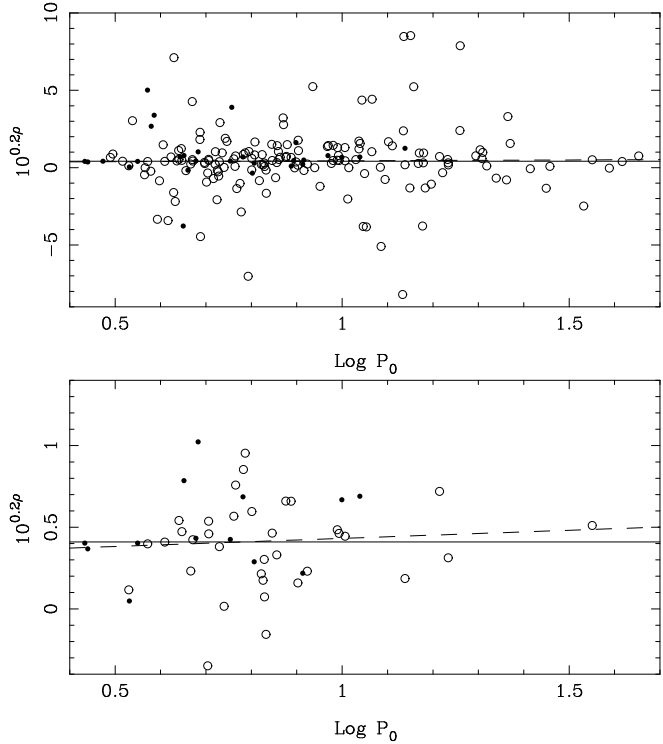,width=8.8cm}}
\caption[]{10$^{0.2 \rho}$ versus $\log P$ based on $VI$ colours for
a slope of $-3.05$. In the top panel all 191 stars are plotted. The
overtones are marked by the filled symbols. In the bottom panel only
the 48 stars with an individual weight $>5$ are shown. The solid lines
represent the weighted mean of 10$^{0.2 \rho}$ for the respective
samples. The dashed lines represent the weighted least-squares fit to
the data for the respective samples.}
\end{figure}

\subsection{Zero point of the $M_{\rm K}-\log P$-relation}

Table~3 lists the results for the zero point of the $M_{\rm K}-\log
P$-relation and the zero point for the $PL$-relations in $V$ and $I$
for the same samples. Unless noted otherwise we have used a slope of
$\delta = -3.27$ from Gieren et al. (1998). For the whole sample we
find a zero point of $-2.61 \pm 0.17$, but note that the corresponding
zero points in the $V$ and $I$ band of this sub-sample are too bright
by about 0.11 magnitude compared to the full samples in $V$ and $I$,
and so the true zero point is probably closer to $-2.50$. These
off-sets indicate biases due to the small number of available
measurements in the NIR. This is especially clear when further
sub-divisions into smaller samples are made, like a division in period
(solutions 9-13).  Especially in one bin (solution 10) the result
depends very much on one star which was taken out (solution 11). After
shifting the zero points in $K$ by an amount such as to make the zero
points in $V$ all equal there is no evidence for a dependence of the
zero point on period.

\begin{table*}
\caption[]{Values for the zero point from $JK$ photometry.}
\begin{tabular}{rrcrccl} \hline
Solution& N  &   Zero point & Total  & Zero point &  Zero point & Remarks \\
        &    &    in $K$    & Weight &  in $V$    &   in $I$    &      \\ \hline
1 & 63 & -2.607 $\pm$ 0.169 & 1821.0 & -1.52 $\pm$ 0.17 & -2.00 $\pm$
        0.17 & All stars \\
2 & 56 & -2.645 $\pm$ 0.192 & 1455.6 & -1.56 $\pm$ 0.19 & -2.02 $\pm$ 0.20 & All fundamental modes \\
3 &  7 & -2.462 $\pm$ 0.353 &  365.3 & -1.41 $\pm$ 0.35 & -1.94 $\pm$ 0.36 & All overtones \\
4 &  1 & -2.692 $\pm$ 0.398 &  355.5 & -1.52 $\pm$ 0.39 & -2.09 $\pm$ 0.41 & $\delta$ Cep \\
5 &  6 & -2.602 $\pm$ 0.227 & 1006.9 & -1.50 $\pm$ 0.22 & -1.97 $\pm$
        0.23 & $V_{\rm obs} < 5.5$ \\
6 & 52 & -2.476 $\pm$ 0.161 & 1775.6 & -1.39 $\pm$ 0.16 & -1.87 $\pm$
        0.16 & $\pi >0$, all stars \\
7 & 45 & -2.479 $\pm$ 0.181 & 1410.2 & -1.39 $\pm$ 0.18 & -1.85 $\pm$
        0.19 & $\pi >0$, fundamental modes \\
8 &  7 & -2.462 $\pm$ 0.353 &  365.4 & -1.41 $\pm$ 0.35 & -1.94 $\pm$
        0.36 & $\pi >0$, overtones \\
9 &  8 & -2.601 $\pm$ 0.392 &  337.2 & -1.59 $\pm$ 0.39 & -2.08 $\pm$ 0.40 & $\log P < 0.64$ \\
10& 17 & -3.002 $\pm$ 0.445 &  377.9 & -1.91 $\pm$ 0.45 & -2.38 $\pm$
        0.45 & $0.64 \le \log P < 0.85$, no $\delta$ Cep \\
11& 16 & -2.746 $\pm$ 0.400 &  370.4 & -1.61 $\pm$ 0.40 & -2.11 $\pm$
        0.40 & $0.64 \le \log P < 0.85$, no $\delta$ Cep, no V496 Aql \\
12&  8 & -2.833 $\pm$ 0.494 &  262.1 & -1.70 $\pm$ 0.49 & -2.24 $\pm$ 0.50 & $0.85 \le \log P < 0.99$ \\
13& 29 & -2.195 $\pm$ 0.270 &  485.3 & -1.18 $\pm$ 0.27 & -1.50 $\pm$
        0.27 & $\log P \ge 0.99$ \\
14&  3 & -2.635 $\pm$ 0.271 &  725.1 & -1.49 $\pm$ 0.27 & -2.01 $\pm$
        0.28 & ${\pi}_{\rm phot} >$ 2.6 mas, fundamental mode\\
15& 63 & -2.599 $\pm$ 0.169 & 1807.9 & -1.52 $\pm$ 0.17 & -2.00 $\pm$ 0.17 & as (1), $K$ larger by 0.005 \\
16& 63 & -2.611 $\pm$ 0.169 & 1827.7 & -1.52 $\pm$ 0.17 & -2.00 $\pm$ 0.17 & as (1), $J-K$ larger by 0.007 \\
17& 63 & -2.604 $\pm$ 0.169 & 1816.8 & -1.52 $\pm$ 0.17 & -2.00 $\pm$ 0.17 & as (1), $(J-K)_0$ larger by 0.004 \\
18& 63 & -2.612 $\pm$ 0.169 & 1830.9 & -1.52 $\pm$ 0.17 & -2.00 $\pm$ 0.17 & as (1), $R(K)$ larger by 0.097 \\
19& 62 & -2.800 $\pm$ 0.177 & 1979.3 & -2.06 $\pm$ 0.18 & -2.60 $\pm$ 0.18 & $\delta = -3.05$, $\log P > 0.5$ \\
20& 62 & -2.321 $\pm$ 0.178 & 1262.8 & -1.31 $\pm$ 0.18 & -1.74 $\pm$ 0.18 & $\delta = -3.60$, $\log P > 0.5$ \\
\hline
\end{tabular}
\end{table*}

\subsection{Zero point of the $M_{\rm W}-\log P$-relation}

Table~4 lists the results for the zero point of the $M_{\rm W}-\log
P$-relation. We have used a slope of $\delta = -3.411$ (Tanvir
1999). As before the sample is divided into period bins. After
shifting the zero points in $W$ by an amount such as to make the zero
points in $V$ all equal (this value can be taken from the
corresponding solutions in Table~2) there is no evidence for a
dependence of the zero point on period.

\begin{table*}
\caption[]{Values for the zero point from the Wesenheit-index.}
\begin{tabular}{rrcrl} \hline
Solution& N  &   Zero point & Total   & Remarks \\
        &    &    in $W$    & Weight  &      \\ \hline
1 & 191 & -2.557 $\pm$ 0.104 & 4554.1  & All stars \\
2 & 163 & -2.622 $\pm$ 0.156 & 2167.2  & All fundamental modes \\
3 &  28 & -2.500 $\pm$ 0.141 & 2387.0  & All overtones \\
4 &   1 & -2.528 $\pm$ 0.176 & 1555.8  & Polaris \\
5 & 139 & -2.424 $\pm$ 0.100 & 4431.9  & $\pi > 0$, all stars \\
6 &  35 & -2.626 $\pm$ 0.267 &  743.3  & $\log P < 0.66$, no Polaris \\
7 &  36 & -2.704 $\pm$ 0.277 &  737.2  & $0.66 \ge \log P < 0.78$ \\
8 &  44 & -2.800 $\pm$ 0.283 &  771.4  & $0.78 \ge \log P < 0.92$ \\
9 &  75 & -2.204 $\pm$ 0.219 &  746.5  & $\log P \ge 0.92$ \\
10& 185 & -2.558 $\pm$ 0.136 & 2676.9  & $\log P \ge 0.50$ \\
11&  12 & -2.585 $\pm$ 0.127 & 3118.8  & ${\pi}_{\rm phot} > 1.8$ mas \\
12&  10 & -2.569 $\pm$ 0.134 & 2808.9  & ${\pi}_{\rm phot} > 1.8$ mas,
        $\log P \ge 0.50$ \\
13& 191 & -2.581 $\pm$ 0.104 & 4656.6  & As (1), reddening coefficient
        of 2.45 in Eq.~(11) \\
14& 191 & -2.564 $\pm$ 0.104 & 4584.0  & As (1), $V$ larger by 0.005 \\
15& 191 & -2.545 $\pm$ 0.104 & 4503.7  & As (1), $I$ larger by 0.005 \\
\hline
\end{tabular}
\end{table*}

\section{Summary of the results}

\subsection{On the use of longer wavelengths}

One argument to consider $PL$-relations in $I$ and $K$ besides the
traditional relation in $V$, is because of the smaller extinction at
longer wavelengths. We have calculated the zero points for the whole
sample considering a systematic shift for all stars of 0.005 mag in
$V$ (respectively $I$ and $K$), 0.007 in $B-V$ (resp. $V-I$, $J-K$), a
shift in the period-colour relations (Eqs.~(4), (8), (12)) equal to
1/10-th of the quoted dispersions, and a shift in the selective
reddening (Eqs.~(5), (10), (14)) equal to the quoted dispersion. The
results are listed in Table~1 (solutions 48-51), Table~2 (solutions
23-26), and Table~3 (solutions 15-18). Adding the differences between
these zero points and that for the default case in quadrature, one
arrives at estimated uncertainties due to errors in the photometry,
reddening and period-colour relations of 0.038 mag in $V$, 0.032 mag
in $I$, and 0.011 mag in $K$. This illustrates that the $K$-band
$PL$-relation is indeed the least sensitive to these effects. For the
Wesenheit-index the error due to the photometry and reddening
coefficients amounts to 0.028 mag (solutions 13-15 in Table~4). \\

Unfortunately, the advantages of using the infrared, like the
intrinsically tighter $PL$-relation and the insensitivity to
reddening, are countered by the fact that so few stars have been
measured in the NIR so far. Of the approximately 48 stars with in 1
kpc, $BV$ photometry is available for all of them, $VI$ for 41 of
them, but $JHK$ data (of sufficient quality) for only 24. It is
estimated that determining the intensity-mean NIR magnitudes of the
remaining 24 stars alone would bring the scatter in the zero point
determination in the $K$-band down from about 0.17 to less than 0.1 mag. \\

\subsection{On the slopes of Galactic $PL$-relations}

Another important issue concerns the slopes in the respective Galactic
$PL$-relations. Common practice is to adopt the slope determined for
\C\ in the LMC, but the slope could be different for Galactic \C.  In
Table~5 we have collected slopes of the $PL$-relations from the
literature for \C\ in the Galaxy, LMC and SMC, in $V,I,K$, both
observationally determined and from two recent theoretical papers
(Alibert et al. 1999, Bono et al. 1999 and private communication). 
Table~5 includes ths slopes in $V$ and $I$ by Madore \& Freedman (1991)
used by the {\sc hst} $H_0$ Key Project (see for example Gibson et
al. 1999).  From the results of Gieren et al. (1998) on the \C\ in the
LMC, we have calculated the error in the slope using the data they
kindly provided (their Tables 8 and 9). In addition we have calculated
the slope and zero point for their data set but using cut-offs in
period, for reasons that will be explained later.  \\

There are several interesting features to be noted about the slopes.
Observationally, the slopes in the $V$ and $I$ $PL$-relations in the
LMC are very well established, respectively, about $-2.81$ and
$-3.05$, and these are the default slopes adopted in the present study,
with errors of about 0.08 and 0.06. For $M_{\rm K}-\log P$-relation
there are fewer observational data available but the slope in Gieren
et al. (1998) is determined very accurately and is in reasonable
agreement with the result of Madore \& Freedman (1991). \\

One fact needs to be mentioned however, and that is that the period
distribution of the calibrating \C\ in the LMC is very different from
that of the Galactic \C\ in the \HP\ catalog. In Table 8 in Gieren et
al. (1998) there are 53 LMC \C\ listed with $V,I$ photometry that
define their $PL$-relation.  Nine have $\log P \le 0.555$, then there
is a gap, and 42 have $\log P \ge 0.846$. In $JHK$ (their Table 9),
there are 59 \C, including 5 that have $\log P \le 0.680$, then there
is a gap, and 54 have $\log P \ge 0.834$.  The same dichotomy of \C\
in period can be seen from Tanvir (1999). Note that in more distant
Galaxies than the Magellanic Clouds, due to observational bias, most
known Cepheids have periods longer than 10 days. We therefore have
included in Table~5 the slopes in $V,I,K$ based on the data in Gieren
et al. (1998), but have divided the sample according to period. The
slopes for the long period sub-sample differ only slightly from that
for the whole sample but are characterised by slightly larger
errors. The slopes for the short-period sample have larger errors
because of the small number of stars involved but nevertheless are
systematically steeper at the 1$\sigma$ level in all three
colours. For comparison, in our sample only 107 of the 236 \C\ have
$\log P \ge 0.846$, and the zero point of this sample differs at the
1.4$\sigma$ level from that of the whole sample (solution 52 in
Table~1). This is yet another indication that for the default slope of
$-2.81$ the zero point may depend on the period range chosen. For a slope
of $\delta = -2.22$ (compare solutions 38, 53) this is not the case. \\

A second issue concerns the predictions of the theoretical models and the
comparison with observations. In $V,K$ for the LMC, and in $I$ for the
SMC the models of Bono et al. (1999) are in very good agreement with
the observed slope, in $I$ for the LMC and $V$ for the SMC the
agreement is poor. In $I, K$ for the LMC, the models of Alibert et
al. (1999) are in good agreement with the observed slopes; in $V$ the
agreement for the LMC and $V,I$ for the SMC is fair. However, the
prediction both models make for the Galaxy are very different. Gieren
et al. (1998) have derived $PL$-relations for Galactic \C\ using the
surface brightness technique. The slopes they find in all three bands
are {\it steeper} than the corresponding slopes in the LMC at the 2-3
$\sigma$ level. In their paper they ascribe this to small number
statistics, and in the end assume the LMC slopes to hold for the
Galactic \C\ as well, also, they add, because the slopes for the LMC
\C\ are better determined.  The models of Alibert et al. (1999)
predict slopes for the Galactic $PL$-relations that are {\it steeper}
than the observed ones in the LMC in $V,I,K$ as well, although by a
small amount only (the main conclusion of their paper is actually that
the slope and zero point of the $PL$-relations do not depend on
metallicity). By contrast, the Bono et al. (1999) paper predicts
slopes that are significantly {\it shallower} in the Galaxy compared
to the LMC especially in $V,I$ but also in $K$. In addition, the Bono
et al. models actually predict a non-linear $PL$-relation in $V$, for
all metallicities considered.  Furthermore, Alibert et al. mention
that a change of slope in $PL$-relations is expected at short periods
due to the reduction of the blue loop during core He burning, and that
this change of slope occurs near $\log P = 0.2$ for $Z$ = 0.004, $\log
P = 0.35$ for $Z$ = 0.01, and $\log P = 0.5$ for $Z$ = 0.02. Such a
change of slope (in the sense that the slope of the $PL$-relation is
steeper for periods below this limit) was recently observed for SMC
\C\ (Bauer et al. 1999) with the break occurring near $\log P = 0.3$,
near the predicted value. \\

\begin{table*}
\caption[]{Slopes of the $PL$-relations.}
\begin{tabular}{ccll} \hline
Slope         & Colour & System & Reference \\ \hline
$-3.037 \pm 0.138$ & V & GAL & Gieren et al. (1998)\\
 $-2.22 \pm 0.04$  & V & 0.02& Bono et al. (1999); non-linear, $\log P<1.4$\\
$-2.905$           & V & 0.02& Alibert et al. (1999)  \\

$-2.810 \pm 0.082$ & V & LMC & Tanvir (1999) \\
$-2.769 \pm 0.073$ & V & LMC & Gieren et al. (1998) \\
$-2.820 \pm 0.118$ & V & LMC & this work; 44 stars with $\log P >
0.845$ from Gieren et al. (1998)\\
 $-3.54 \pm 0.68$  & V & LMC & this work;  9 stars with $\log P < 0.845$ from Gieren et al. (1998)\\
 $-2.88 \pm 0.20$  & V & LMC & Madore \& Freedman (1991)\\
 $-2.81 \pm 0.06$  & V & LMC & Caldwell \& Laney (1991)\\
 $-2.79 \pm 0.06$  & V & 0.008& Bono et al. (1999); non-linear, $\log P<1.4$\\
$-2.951$           & V & 0.01& Alibert et al. (1999)  \\

$-2.63 \pm 0.08$   & V & SMC & Caldwell \& Laney (1991), \\
$-3.04 \pm 0.04$   & V & 0.004& Bono et al. (1999); non-linear, $\log P<1.4$\\
$-2.939$           & V & 0.004& Alibert et al. (1999)  \\

$-3.329 \pm 0.132$ & I & GAL & Gieren et al. (1998)\\
 $-2.35 \pm 0.08$  & I & 0.02& Bono et al. (1999)  \\
$-3.102$           & I & 0.02& Alibert et al. (1999)\\

$-3.078 \pm 0.059$ & I & LMC & Tanvir (1999) \\
$-3.041 \pm 0.054$ & I & LMC & Gieren et al. (1998)\\
$-3.084 \pm 0.088$ & I & LMC & this work; 44 stars with $\log P > 0.845$ from Gieren et al. (1998)\\
 $-3.39 \pm 0.39$  & I & LMC & this work;  9 stars with $\log P < 0.845$ from Gieren et al. (1998)\\
 $-3.14 \pm 0.17$  & I & LMC & Madore \& Freedman (1991)\\
 $-3.01 \pm 0.05$  & I & LMC & Caldwell \& Laney (1991) \\
 $-2.63 \pm 0.08$  & I & 0.008& Bono et al. (1999)  \\
 $-3.140$          & I & 0.01& Alibert et al. (1999) \\

 $-2.92 \pm 0.07$  & I & SMC & Caldwell \& Laney (1991)  \\
 $-2.73 \pm 0.10$  & I & 0.004& Bono et al. (1999)  \\
 $-3.124$          & I & 0.004& Alibert et al. (1999) \\

$-3.598 \pm 0.114$ & K & GAL & Gieren et al. (1998) \\
 $-3.03 \pm 0.07$  & K & 0.02 & Bono et al. (1999)  \\
 $-3.367$          & K & 0.02 & Alibert et al. (1999)\\

$-3.267 \pm 0.041$ & K & LMC & Gieren et al. (1998) \\
$-3.304 \pm 0.052$ & K & LMC & this work; 54 stars with $\log P > 0.833$ from Gieren et al. (1998)\\
 $-3.37 \pm 0.39$  & K & LMC & this work;  5 stars with $\log P < 0.833$ from Gieren et al. (1998)\\
 $-3.42 \pm 0.09$  & K & LMC & Madore \& Freedman (1991) \\
 $-3.19 \pm 0.09$  & K & 0.008 & Bono et al. (1999) \\
 $-3.395$          & K & 0.01& Alibert et al. (1999) \\

 $-3.27 \pm 0.09$  & K & 0.004 & Bono et al. (1999)  \\
 $-3.369$          & K & 0.004 & Alibert et al. (1999) \\

$-3.411 \pm 0.036$ & Wesenheit & LMC & Tanvir (1999) \\


\hline
\end{tabular}
\end{table*}

\subsection{Consistency between individual distances based on 
different colours}

For a given slope and zero point of the $PL$-relation one can
calculate the photometric distance, and since we have determined the
zero point for three photometric band we can intercompare photometric
distances to individual \C. This is illustrated in Fig.~6, for the
default slopes of the $PL$-relations. The zero point of the $M_{\rm
V}-\log P$ relation is fixed at $-1.411$ (solution 10, Table~1). The
zero point of the $M_{\rm I}-\log P$ relation is determined to give a
mean difference in $(m-M)_{\rm 0,I} - (m-M)_{\rm 0,V}$ of zero, and is
found to be $-1.918$ (top panel Fig.~6). The rms dispersion is 0.14
mag.  Similarly, to create the bottom panel, the zero point of the
$M_{\rm K}-\log P$ relation was determined to give a mean difference
in $(m-M)_{\rm 0,K} - (m-M)_{\rm 0,I}$ of zero, and is found to be
$-2.600$, with an rms dispersion is 0.10. The values for the zero
points in $I$ and $K$ derived in this way differ only slightly from
the solutions 1 in Tables 2 and 3.

Interestingly, least-square fitting shows that the slopes of the
relations are not unity, but 0.980 $\pm$ 0.007 (top panel), and 0.965
$\pm$ 0.007 (bottom panel).  Using the same procedure, but adopting
steeper slopes of $-3.04$, $-3.33$, $-3.60$ (see Sect.~8 for the
reason of these choices) in, respectively, the $M_{\rm V}-\log P$,
$M_{\rm I}-\log P$ and $M_{\rm K}-\log P$ relations and fixing the
zero point of the $M_{\rm V}-\log P$ relation at $-1.234$ we find in
the same way the zero points of the $M_{\rm I}-\log P$ and $M_{\rm
K}-\log P$-relations to be $-1.696$ and $-2.328$. The slopes are 0.983
$\pm$ 0.007 and 0.971 $\pm$ 0.007.

Using the same procedure, but adopting shallower slopes of $-2.22$,
$-2.35$, $-3.05$ in, respectively, the $M_{\rm V}-\log P$, $M_{\rm
I}-\log P$ and $M_{\rm K}-\log P$ relations and fixing the zero point
of the $M_{\rm V}-\log P$ relation at $-1.885$ (solution 38 in Table~1)
we find the zero point of the $M_{\rm I}-\log P$ and $M_{\rm K}-\log
P$-relations to be $-2.491$ and $-2.692$. The slopes are 0.974 $\pm$
0.007 and 1.010 $\pm$ 0.010.

The conclusion is the the photometric distances based on different
$PL$-relations are consistent with each other at a level of 0.10-0.14
mag, similar to the uncertainties in the individually derived zero
points in $V,I,K$ for the full sample. The data does not allow to
discriminate between different choices of the slopes of the $PL$-relations.

\begin{figure}
\centerline{\psfig{figure=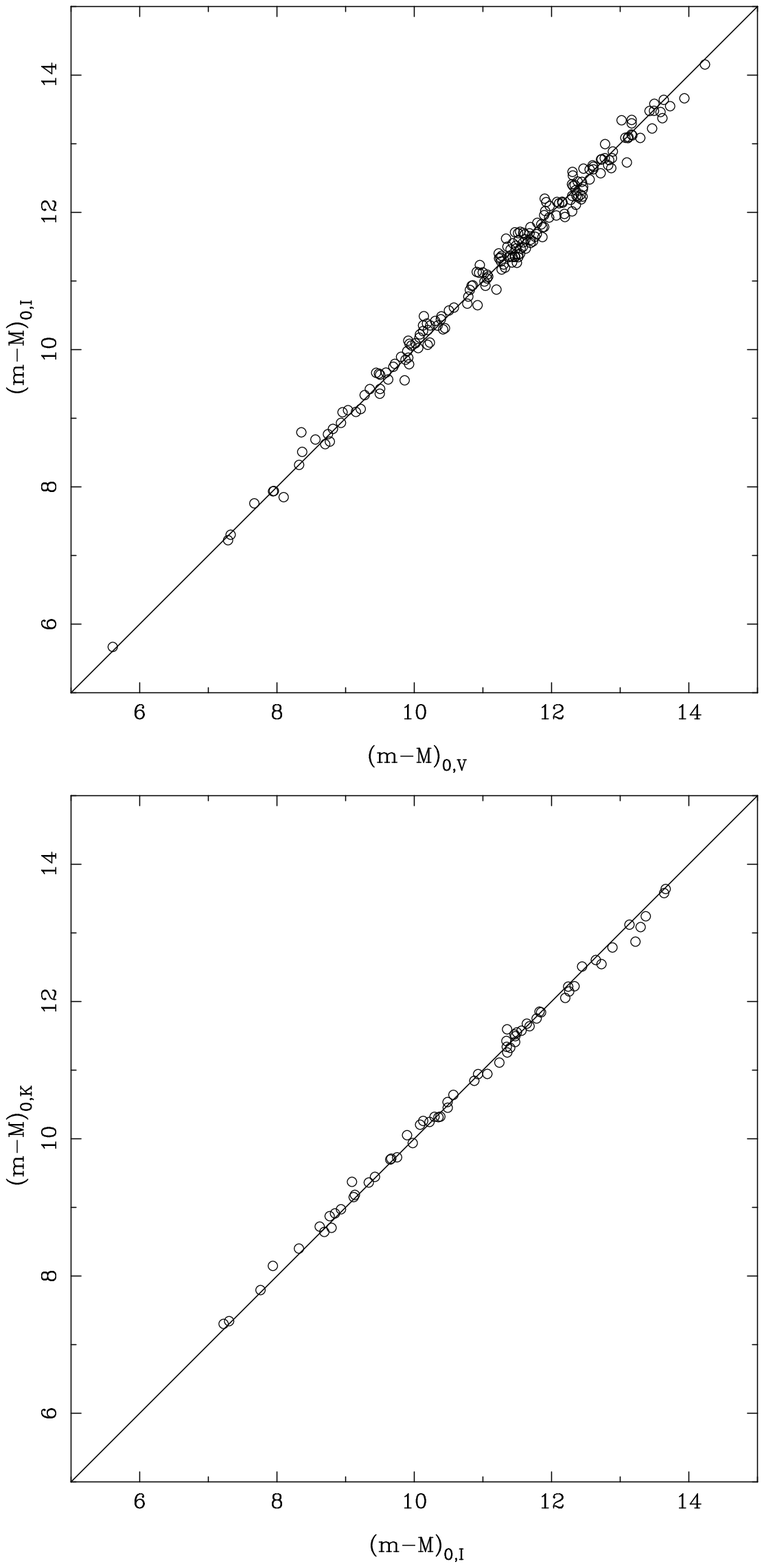,width=8.8cm}}
\caption[]{Comparison of the distance moduli based on the
$PL$-relations in $V$ and $I$ (top panel), assuming zero points of
respectively $-1.411$ and $-1.918$, and the default slopes, and $I$
and $K$ (bottom panel) for zero points of $-1.918$ and $-2.600$, and
the default slopes. The solid line is the one-to-one relation,
although a least-square fit indicates that the best fitting slope
differs slightly from unity (see text).
}
\end{figure}

\section{A volume complete sample}

The question remains how representative the \HP\ Cepheid sample is. As
with all stars that made it in the \HP\ catalog, they were proposed by
Principal Investigators in solicited proposals. If proposers preferred
their `favorite' objects to be observed, there may be a `Human' bias,
which is impossible to correct for. For example, at some point in the
selection process of the \HP\ Input Catalog the number of \C\ with
magnitudes between 9.5 and 11 were `tuned-up' (ESA 1989, page
80). Also, the distribution in pulsation period of the about 72 \C\
that were proposed to be observed but did not end up in the \HP\ Input
Catalog is somewhat different from the 270 that did make it (ESA 1989,
page 96).

In view of this it may be instructive to try to construct a volume
complete sample.  The \HP\ Input Catalog (ESA 1989, page 94) mentions
that {\it all} 55 \C\ within 1 kpc have been included. This depends of
course on the adopted slope and zero point to calculate the
photometric parallax. The F95 database lists 51 stars within 1 kpc for
their adopted relation of $M_{\rm V} = -2.902 \, \log P -1.203$.

In Fig.~7 the  cumulative (photometric) parallax distribution is
plotted for the 236 stars in the sample, with distances from F95, or
calculated in the same way for the stars not listed there. Also
indicated is a line with slope $-2$ as expected for a disk population
(with $N(r)dr \sim r$ it follows that $N(\pi)d\pi \sim {\pi}^{-3}$,
and so the cumulative distribution is proportional to
${\pi}^{-2}$). From this figure it is confirmed that the \HP\ catalog
is volume complete to approximately 1 kpc.

On the other hand, three new \C\ were discovered with \HP. For the same
slope and zero point as used in F95 the photometric parallaxes of CK
Cam, V898 Cen, V411 Lac are 1.77, 0.73 and 1.17 mas, respectively.


As \HP\ discovered 2 \C\ that are closer than 1 kpc, it implies that
the present sample may now be indeed complete down to 1 mas, but is
possibly only complete for stars with a parallax \more 1.8 mas as no
new \C\ closer than this limit have been discovered.

In Tables~1-4 we include solutions for volume complete samples, or, in
other words, complete in photometric parallax. Since this depends on
the zero point itself, this is an iterative process. For the $M_{\rm
V}-\log P$-relation we give the results in Table~1 for a cut-off at 1
and 1.8 mas. In solutions 39, 41, 43 we list the zero point for an
input zero point of $-1.411$ (solution 10) for the whole sample, the
fundamental pulsators, and the overtones, respectively.  The next line
(solutions 40, 42, 44) lists the final result after iteration on the
zero point. Solutions 45-47 give the final results for a higher
cut-off in photometric parallax. In most cases the zero points are
slightly brighter than the corresponding solutions 10, 11, 12. This is
a bit surprising because of the Malmquist bias one expects the volume
complete sample to be dimmer. The expected Malmquist bias for a disk
population with an intrinsic spread of ${\sigma}_{\rm H}$ = 0.10
around the mean is about 0.009 mag (Stobie et al. 1989). This is much
smaller than the typical error estimate in the zero points, and so the
fact that the volume complete sample is brighter probably reflects the
slightly different nature of that sample.  For example, the mean value
of $\log P_0$ for the whole sample is 0.86, while that for the volume
complete sample is 0.77. As we noticed earlier, since the shortest
period bins give brighter zero points this could be an explanation. \\

For the $M_{\rm I}-\log P$-relation the construction of a volume
complete sample is slightly more complicated due to the fact that for
some of the 47 stars with $BV$ photometry and a photometric parallax
$>1$ mas no $I$-band photometry is available. This implies that one has to
apply a more stringent criterion to obtain a volume complete sample of
stars that also have $I$-band photometry. This turns out to be a
photometric parallax limit of 1.8 mas, and solution 23 lists the
result. The zero point is slightly brighter, and this again may be due
to the fact that the average $\log P_0$ value is smaller (0.74) than
for the whole sample (0.90). \\

In the $K$-band no meaningful volume complete sample can be
constructed.  As there is no NIR data available for Polaris, any
volume complete sample could be constructed for fundamental pulsators
only. In any case, even then, a volume complete sample of stars with
NIR data would have a cut-off at 2.6 mas, and would only include 3
stars. For completeness, it has been included in Table~3 (sol. 14). \\

The conclusion is that the zero point of the $PL$-relation based on a
volume complete sample are within $1\sigma$ of the results for the
full sample. Surprisingly, the zero points are brighter, contrary to
one would expect from Malmquist bias. However, the Malmquist
correction is expected to be small (see the next section for an
estimate based on numerical simulations), and certainly much smaller
than the error in the zero point, so that this effect is due to the
slightly different nature of the volume complete samples compared to
the full samples.

\begin{figure}
\centerline{\psfig{figure=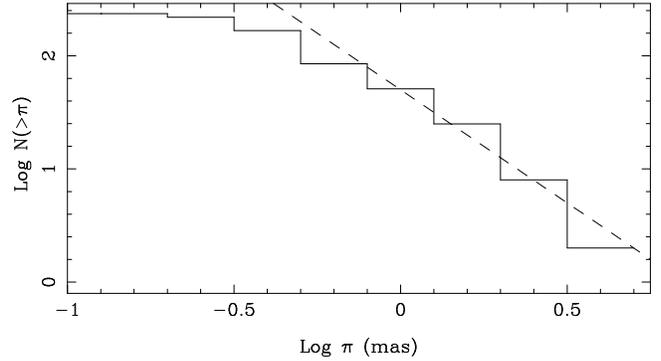,width=8.8cm,angle=-90}}
\caption[]{Cumulative (photometric) parallax distribution of the 236
stars in the sample with distances from F95 who calculated them using
$M_{\rm V} = -2.902 \, \log P -1.203$. The dashed line has a slope of
$-2$, as expected for a disk population. It shows that \HP\ is volume
complete down to a distance of about 1 kpc.}
\end{figure}

\section{Properties of the \HP\ \C, and numerical simulations}

In this section we describe a numerical model to first of all construct
synthetic samples of stars that fit some observed properties of the
\HP\ \C. Second, this model is used to investigate the numerical bias
involved in applying the ``reduced parallax'' method. \\

The observed distributions of the fundamental period, the $V$
magnitude, de-reddened $(B-V)_0$, colour excess, and absolute distance
to the Galactic plane are plotted in Fig.~8 for the 236 \C\ in our
sample (the dotted lines). For the overtones, the fundamental period
was calculated following Eqs.~(6-7). $(B-V)_0$ follows from Eq.~(4),
and $E(B-V)$ from the observed value of $(B-V)$ minus $(B-V)_0$. The
absolute distance to the Galactic plane is calculated from the
galactic latitude combined with the de-reddened $V$-magnitude and a
photometric distance based on Eq.~(1) with $\delta = -2.81$ and $\rho
= -1.43$. The distributions are compared to simulations described now.

We have devised a numerical code to simulate the distributions
described above.  The input period distribution is (assumed ad-hoc to
be) a Gaussian in $\log P$ with mean $X_{\rm P}$, and spread $X_{\rm
\sigma P}$. Based on the observed properties of the sample only $\log
P$ values between 0.43 and 1.66 are allowed. A random number is drawn
and a $\log P$ selected following this distribution. The values of
$X_{\rm P}$ and $X_{\rm \sigma P}$ directly influence the resulting
distribution and so are easily determined. We find that values of
$X_{\rm P}$ between 0.65 and 0.75 and of $X_{\rm \sigma P}$ between
0.2 and 0.25 give acceptable fits.

The galactic distribution of \C\ is assumed to be a double exponential
disk with a scale height $H$ in the $z$-direction (the coordinate
perpendicular to the galactic plane), and a scale height $R_{\rm GC}$
in the galacto-centric direction. The coordinate system used is
cylindrical coordinates centered on the Galactic Centre. Three random
numbers are drawn to select the distance to the Galactic plane, the
distance to the Galactic centre and a random angle $\phi$ between 0
and 2$\pi$ in the Galactic plane centered on the Galactic centre. From
this the distance $d$ to the Sun is calculated.  Based on the observed
photometric parallaxes of the sample only stars closer than 7800 pc to
the Sun are allowed. We find no evidence for a gradient in the number
of \C\ with galacto-centric radius, in other words $R_{\rm GC} =
\infty$ gives good fits, probably indicating that the volume sampled
is too small to detect such a gradient, or that the distribution of
\C\ is at least equally determined by another factor, for example the
location of the spiral arms. The value of $H$ is directly determined
by the distribution of the number of stars as a function of $z$, and
is found to be between 60 and 80 pc. This is consistent with the scale
height of a relative massive population of stars, as the \C\ are.

The value of $M_{\rm V}$ and $(B-V)_0$ are correlated in the sense
that brighter \C\ are also bluer for a given period. Contrary to other
studies we do not assume a Gaussian spread around the $M_{\rm V}- \log
P$ and $(B-V)_0 - \log P$ relations, but instead a `box' like
distribution which is more physical because of the finite width of the
instability strip. However, this does assume that the instability
strip is uniformly filled (for all periods).

A single uniform random number, $Rn$, is drawn and then
\begin{equation}
 M_{\rm V} = -2.81 \, \log P - 1.43 + \; (-0.42 + 0.84 \times {\rm Rn})
\end{equation}
and
\begin{equation}
  (B-V)_0 \;= 0.416 \, \log P + 0.314 + \; (-0.15 + 0.3 \times {\rm Rn})
\end{equation}
are calculated. The full width of the instability strip in $M_{\rm V}$
is taken to be 0.84 magnitude (derived from plots in Gieren et
al. 1998, Tanvir 1999), and that of the $(B-V)_0$ relation to be 0.3 mag.
(Laney \& Stobie 1994).

\noindent
The reddening is calculated from:
\begin{equation}
 A_{\rm V} = 0.09 \, \frac{1 - \exp \left( -0.0111 \; d\; \sin b
 \right)}{\sin b}
\end{equation}
where $b$ is the absolute value of the galactic latitude. The colour
excess is then calculated using Eq.~(5). The factor in front was varied to
fit the $E(B-V)$ distribution. We also tried the reddening model of
Arenou et al. (1992), but found that it could not fit the $E(B-V)$
distribution.

The simulated `observed' visual magnitude is calculated from $V =
M_{\rm V}+ 5 \; \log d - 5 + A_{\rm V}$. Then a term is added
simulating the uncertainty in the observed $V$, which is described
by a Gaussian distribution with a dispersion of 0.005 mag.

\HP\ was complete only  down to about $V = 7$ and the following
function was used to determine if a star was `observed' or not. A
random number ($Rn$) was drawn to calculate:
\begin{equation}
 V_{\rm lim} = 7.0 - \frac{\log (1.0 - {\rm Rn} )}{C}
\end{equation}
with $C$ empirically determined to be:
\begin{equation}
 C = 0.107 -0.030\; y -0.00482 \; y^2 +0.00361 \; y^3
\end{equation}
with $y = (V-7)$. A star is `observed' if $V < V_{\rm lim}$.

The simulation is continued until 25 sets of 236 stars fulfill all
criteria. Typically 100~000 stars needed to be drawn to arrive at
this. The simulated distributions are depicted in Fig.~8 using the
solid lines (normalised to the observed number of 236 stars). Typical
parameters $X_{\rm P} = 0.70$, $X_{\rm \sigma P} = 0.25$, and $H = 70$
pc have been used. The overall fit is good. \\

This numerical code for the quoted parameters provides us with a tool
with which synthetic samples of Cepheids can be generated that obey the
observed distributions.

The main difference with the simulations of L99 is in the conclusion
about the space distribution of \C. They assume a homogeneous 3D
distribution, and consider a box centered on the Sun of 4200 pc on a
side. They justify this because of ``the relative small depth of the
\HP\ survey [with respect to] the depth of the Galactic disk''. This
is not true however. First of all, using any reasonable combination of
$\delta$ and $\rho$ it is clear that \HP\ sampled \C\ to much greater
distances than 2.1 kpc, in fact out to 7-8 kpc.  Furthermore, since
the scale height of 3-10 \msol\ main-sequence stars (the progenitors
of the \C\ variables) is of order 100 pc it is clear that \HP\ sampled
to distances much larger than the scale height of the Cepheid
population.  This is directly confirmed from our simulations from
which we derive a scale height of about 70 pc. In other words, the
space distribution of the \C\ is a disk population, not a homogeneous
population.

This might have consequences for the results of the L99 paper
concerning biases which are difficult to judge by us.  Another
consequence is related to the fact that both Malmquist- and LK-bias
depend on the underlying distribution of stars. For example, Oudmaijer
et al. (1998) calculated the LK-bias assuming a homogeneous
distribution of \C. For a disk population the values of both the
Malmquist- and LK-bias are smaller (Stobie et al. 1989, Koen 1992). \\

\begin{figure}
\centerline{\psfig{figure=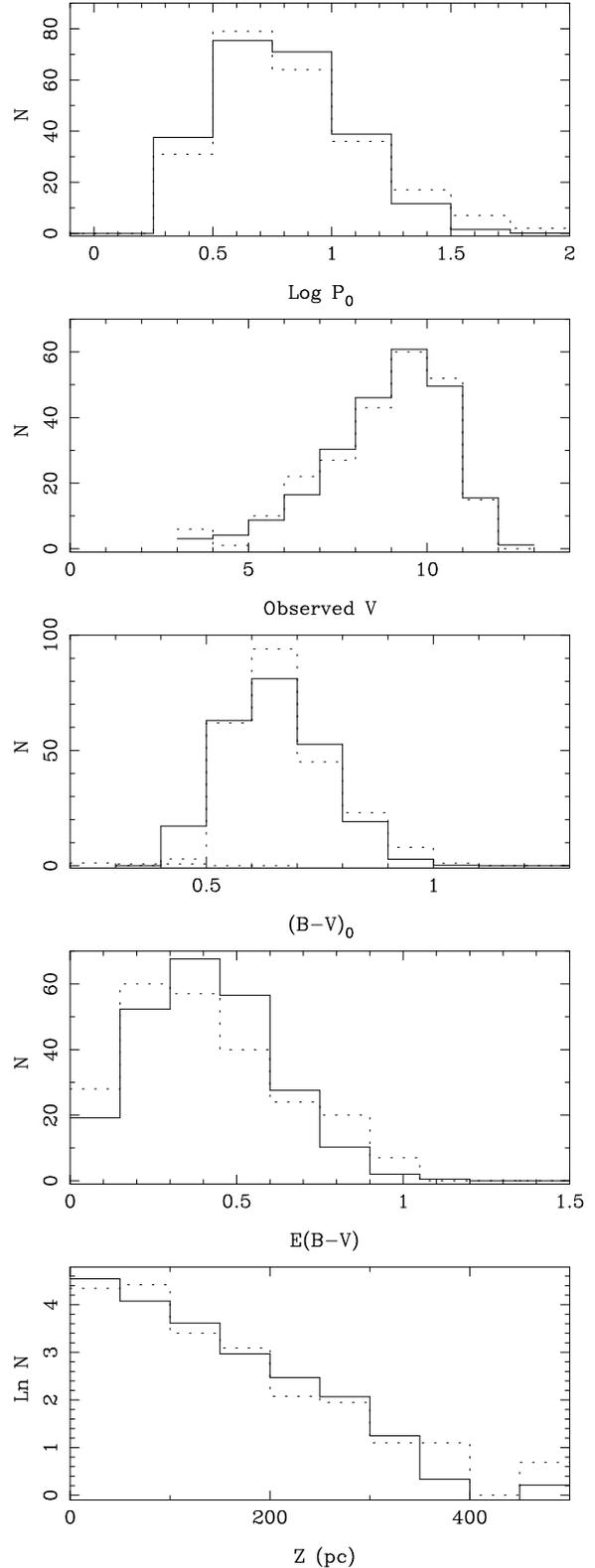,width=7.8cm}}
\caption[]{Observed (dotted line) and numerical simulation (solid
line) of the 236 \C\ in \HP. Harmonic periods have been converted to
the fundamental period. For a description of the model, and the
parameters see Sect.~6.
}
\end{figure}

\noindent
Now, we will discuss whether the zero points derived by the method
outlined in Sect.~4 are subject to bias or not. Similar simulations
were also performed by L99 and Pont (1999).  L99 concluded that zero
points derived for the whole sample in $V$ are too bright by 0.01
mag. Pont (1999) concluded that any bias is less than 0.03 mag.

What remains to be discussed in relation to our numerical model is how
the `observed' parallax and the error in the parallax are calculated.

We define several quantities. First of all a minimum error in the
parallax, calculated from (in mas):
\begin{equation}
\begin{array}{llll}
{\sigma}_{\pi}^{\rm min} & = & 0.45                    & V \le 8 \\
                         & = & 0.45 \times (V/8)^{3.2} & {\rm else} \\
\end{array}
\end{equation} 
Second of all, an error in the error on the parallax, calculated from
(in mas):
\begin{equation}
\begin{array}{llll}
{\sigma}_{\sigma} & = & 0.18                   & V \le 8 \\
     & = & 0.18 - 0.2167\, y + 0.325 \, y^2 - 0.06833 \, y^3 & {\rm else} \\
\end{array}
\end{equation} 
with $y = (V-8)$.
Third, a `mean' error on the parallax, calculated from (in mas):
\begin{equation}
\begin{array}{llll}
{\sigma} & = & 0.80                    & V \le 5 \\
      & = & 0.80 - 0.0280 \, y - 0.00229 \,y^2 + 0.008651 \,y^3 & {\rm else} \\
\end{array}
\end{equation} 
with $y = (V-5)$. All fits have been made with {\sc idl} using the routine
{\sc polyfit} from the full sample of 236 stars. It was verified that
the synthetic samples have the same distribution of parallax and
parallax error compared to the observed sample.

The error on the parallax, ${\sigma}_{\pi}$, was then determined from
$\sigma$ plus a quantity randomly selected from a Gaussian
distribution with dispersion ${\sigma}_{\sigma}$. The result is only
accepted when ${\sigma}_{\pi} > {\sigma}_{\pi}^{\rm min}$ however.
Finally, the `observed' parallax (in mas) is calculated from 1000/$d$,
with $d$ the true distance in pc, plus a quantity randomly selected
from a Gaussian distribution with dispersion ${\sigma}_{\pi}$.

The simulation is continued until 100 sets of 236, 191, or 63 stars
fulfill all criteria, depending whether the simulation relates to
$V$, $I$ or $K$. About 410~000 stars have to been drawn to arrive at
23~600 `observed' stars in $V$. This was done with the parameter set
that best described the observed distributions, as discussed above,
i.e. with $X_{\rm P} = 0.70$, $X_{\rm \sigma P} = 0.25$, and $H = 70$ pc.

In the case the simulation relates to $I$ and $K$ the procedure is as
follows. First the stars are selected according to the procedure
outlined above, that is, selection based on $V$-photometry. For the
stars that fulfill the criteria, the `observed' $I$, $V-I$
(respectively $K$, $J-K$) colours are determined, using the
period-colour and reddening relations outlined in Sect.~4. The {\it
same random number} that was used in Eqs.~(15-16) to calculate the true
$M_{\rm V}$ and $(B-V)$ is used to calculate $M_{\rm I}$ and $(V-I)$,
respectively $M_{\rm K}$ and $(J-K)$. This is to simulate the fact
that if the synthetic star is `observed' to be fainter and bluer than
the mean in $V$, this is also the case at other wavelengths. This
procedure ignores any phase shifts between the light curves at different
colours. For the full-width of the instability strip (cf. Eqs.~ (15-16))
in $M_{\rm I}$ we take 0.70 mag, in $(V-I)$ 0.20 mag, in $M_{\rm K}$
0.60 and in $(J-K)$ 0.16 mag (Gieren et al. 1998).

For every set of 236, 191, or 63 stars we apply the ``reduced
parallax'' method and derive the zero point. From the distribution of
the zero points of the 100 sets, we determine the mean, and the
dispersion.  The results are given in Table~6, where we list the zero
point assumed in the numerical simulation, the zero point retrieved,
the average number of stars in the simulation that fulfilled the
selection criteria, and to which solution the simulation refers
to. From the results we see that the zero point of the volume complete
sample is dimmer than for the whole sample, as expected from Malmquist
bias. In any case, the biases are very small, of order 0.01 mag, or
less, similar to the results found by L99 and Pont (1999). Second, we
confirm the result by Pont (1999) that the errors derived are somewhat
larger compared to the outcome of the ``reduced parallax''
method. However the differences are not so large as compared to the
error bars quoted in FC.

From the simulations we find that the smallest error is for the sample
selected on parallax (as was found for the `real' sample), but that it is
subject to Lutz-Kelker bias.

\begin{table*}
\caption[]{Zero points from numerical simulations.}
\begin{tabular}{ccrcl} \hline
Input    &       Result       & Number & Band & Remarks \\ \hline
$-1.43$  & $-1.425 \pm 0.124$ & 236    & $V$  & all stars, solution 10 \\
$-1.43$  & $-1.418 \pm 0.140$ &  33    & $V$  & volume complete sample, solution 39 \\
$-1.43$  & $-1.431 \pm 0.175$ &  11    & $V$  & $V_{\rm obs} < 5.5$, solution 14 \\
$-1.43$  & $-1.197 \pm 0.114$ & 165    & $V$  & $\pi >0$, solution 16 \\
$-1.43$  & $-1.457 \pm 0.241$ & 215    & $V$  & weight $<$ 11, solution 19 \\
$-1.43$  & $-1.421 \pm 0.289$ &  13.3  & $V$  & 11 $<$ weight $<$ 29, solution 20 \\
$-1.43$  & $-1.443 \pm 0.327$ &   4.8  & $V$  & 29 $<$ weight $<$ 70, solution 21 \\
$-1.43$  & $-1.400 \pm 0.259$ &   2.4  & $V$  & 70 $<$ weight $<$ 500, solution 22 \\
$-1.43$  & $-1.425 \pm 0.130$ & 236    & $V$  & weight $<$ 500, solution 23 \\
$-1.43$  & $-1.416 \pm 0.157$ &  20    & $V$  & 11 $<$ weight $<$ 500, solution 24 \\
         &                    &        &      &         \\
$-1.90$  & $-1.907 \pm 0.143$ & 191    & $I$  & all stars, solution 1 \\
$-1.90$  & $-1.904 \pm 0.206$ &   8.5  & $I$  & volume complete sample, solution 22 \\
$-1.90$  & $-1.664 \pm 0.131$ & 134    & $I$  & $\pi >0$, solution 6 \\
         &                    &        &      &         \\
$-2.60$  & $-2.610 \pm 0.261$ &  63    & $K$  & all stars, solution 1 \\
$-2.60$  & $-2.617 \pm 0.435$ &   1.3  & $K$  & volume complete sample, solution 14 \\
$-2.60$  & $-2.339 \pm 0.234$ &  45    & $K$  & $\pi > 0$, solution 6 \\
\hline
\end{tabular}
\end{table*}

\section{Discussion}

\subsection{The finally adopted zero points}

Based on the results obtained in Sects.~4, 5 and 6, we will present
three sets of solutions, a `traditional' one, and two alternatives.
The traditional one follows FC and L99 closely. The zero point adopted
is the one for the entire sample (which has the lowest error of the
samples that are not selected on observed parallax), adopting the
slope of the $PL$-relation observed for \C\ in the LMC. This would
then be solution 10 from Table~1 ($\rho = -1.41 \pm 0.10$), solution 1
from Table~2 ($\rho = -1.89 \pm 0.11$) and solution 1 from Table~3
corrected for the off-set in the corresponding solution in $V$ and $I$
as discussed in Sect.~4.3 ($\rho = -2.50 \pm 0.17$). These zero points
have to be corrected for Malmquist bias. If the Malmquist bias would
be evaluated in magnitude space, this bias would amount to 0.0092,
0.021 and 0.013 magnitude in $V,I,K$, respectively, for a disk
population and the adopted values for ${\sigma}_{\rm H}$ of,
respectively, 0.10, 0.15 and 0.12 mag (Stobie et al. 1989). However,
if evaluated using the reduced parallax method the Malmquist bias is
smaller (see Oudmaijer et al. 1999), and from the numerical
simulations (Table~6) we find a Malmquist bias of 0.007 in $V$, 0.003
in $I$, and an unphysical negative value in $K$, probably due to the
smaller number of stars involved. For the Malmquist bias we will
assume a (round number) of 0.01 mag in all three bands.

Also, we have increased the errors in the zero points to reflect the
sensitivity to uncertainties in the photometry, reddening and
period-colour relations (see Sect.~5.1). For adopted slopes of
$-2.81$, $-3.05$ and $-3.27$, the finally adopted zero points of the
$PL$-relations in $V$, $I$ and $K$ are, respectively, $\rho = -1.40
\pm 0.11$, $-1.88 \pm 0.12$ and $-2.49 \pm 0.17$. For the
Wesenheit-index, after correcting for Malmquist bias and increasing
the error bar due to the uncertainties described in Sect.~5.1, we
adopt a zero point of $-2.55 \pm 0.11$ for a slope of $-3.411$. \\

The `traditional' solution above has certain advantages, like the fact
that the errors are smallest compared to solutions that do not include
all stars, or, since the slope adopted is the one in the LMC, a
distance determination to the LMC is essentially a comparison of zero
points only (apart from systematic effects). On the other hand,
alternative solutions can be presented which also have merits. These
solutions are based on a volume complete sample (at least in $V$ and
$I$) to avoid Malmquist bias. Although small, its value does depend on
the distribution of stars, and the intrinsic spread in the
$PL$-relation and selecting a volume complete sample avoids Malmquist
bias outright.  In $K$, no volume complete sample could be constructed
and we used the full sample instead.  The alternative solutions take
into account the theoretical prediction that the slope of the
$PL$-relation may change at $\log P \sim 0.5$ for solar metallicities
(Alibert et al. 1999).  Where the two alternative solutions differ are
in the adopted slopes for the Galactic $PL$-relations. One alternative
solution takes into account the, admittedly at the 1$\sigma$ level of
significance, evidence presented in Sects.~4.1-4.2 that the slope of
the Galactic $PL$-relations are shallower than the ones in the LMC,
and in accordance we have adopted the theoretical slopes predicted by
Bono et al. (1999) for solar metallicities. The second alternative
method adopts the slopes in Gieren et al. (1998), who derived
distances from the infrared version of the surface brightness
technique to Galactic \C, which yields steeper slopes than for the
LMC.

These alternative solutions are presented in Tables~1-3 (solutions
54-55, 28-28 and 19-20, respectively).  The zero points in $K$ have to
be corrected for Malmquist bias (+0.01 mag) and for the off-sets in
the corresponding $V$ and $I$ solutions, and the errors in all three
zero points are increased for reasons indicated above.  For the
adopted slopes of $-2.22$, $-2.35$ and $-3.05$, the zero points of the
$PL$-relations in $V$, $I$ and $K$ are, respectively, $\rho = -1.95
\pm 0.12$, $-2.47 \pm 0.15$ and $-2.68 \pm 0.18$. For the adopted
slopes of $-3.04$, $-3.33$ and $-3.60$, the zero points of the
$PL$-relations in $V$, $I$ and $K$ are, respectively, $\rho = -1.30
\pm 0.13$, $-1.75 \pm 0.14$ and $-2.30 \pm 0.18$. \\

\subsection{$PL$-relation of LMC \C}

Before proceeding we have to adopt $PL$-relations for the LMC \C. In
$V$, for a slope of $-2.81$, Caldwell \& Laney (1991) find a zero
point of 17.23 $\pm$ 0.02. Tanvir (1999), for the same slope, gives an
observed zero point of 17.451 $\pm$ 0.043. Adopting a mean reddening
to the LMC of $E(B-V)$ of 0.08 (Caldwell \& Laney 1991) and a ratio of
total-to-selective reddening of $R_{\rm V} = A_{\rm V}/E(B-V) = $ 3.27
(Eq.~(5) for typical colors) we derive a de-reddened zero point of
17.19 $\pm$ 0.04. For the \C\ in the LMC we adopt the weighted mean of
these two values, or a zero point of 17.22 $\pm$ 0.02 in $V$ for a
slope of $-2.81$.

In $I$, for a slope of $-3.041$, Gieren et al. (1998) find a zero
point of 16.74 $\pm$ 0.06 (the error is calculated by us, from their
data). Tanvir (1999), for a slope of $-3.078$, gives an observed zero
point of 16.904 $\pm$ 0.031. Adopting $A_{\rm I} = 0.69 A_{\rm V} =
0.18$ mag for the reddening in $V$ calculated as above, we derive a
zero point of 16.72 $\pm$ 0.03. For the default slope of $-3.05$ we
adopt the weighted mean of these two values, or a de-reddened zero
point in the $I$ band for the \C\ in the LMC of 16.72 $\pm$ 0.02.

In $K$, for a slope of $-3.267$, Gieren et al. (1998) find a zero
point of 16.03 $\pm$ 0.05 (the error is calculated by us from their
data) for the \C\ in the LMC. This value is adopted by us.

For the Wesenheit-index, for a slope of $-3.411$, Tanvir (1999) finds
a zero point of 16.051 $\pm$ 0.017. This value is adopted by us.\\

\subsection{Metallicity correction}

We will now consider the effect of metallicity on the zero point. For
comparison, FC applied a correction of +0.042 mag to the zero point in
the $V$-band, based on Laney \& Stobie (1994). The theoretical models
of Bono et al. (1999), and Alibert et al. (1999) provide
$PL$-relations and from those the difference $\Delta M = M$(Gal) $-
M$(LMC) can be determined which will depend on the photometric band
and period. We have determined this difference for two periods, namely
for $\log P_0 = 0.77$ which we have determined to be the mean period
of the volume complete sample of Galactic \C\ in \HP\, and for $\log
P_0 = 0.47$ which is the mean period of \C\ in the LMC (Alcock et
al. 1999).  The results for $\Delta M$ are listed in Table~7 for the
three photometric bands. This illustrates the difference between the
two theoretical models, for the Alibert et al. (1999) models predict
essentially no dependence on metallicity, while the Bono et al. (1999)
models predict a significant metallicity dependence, which mostly is
in the sense that the metal-rich pulsators are fainter than the
metal-poor ones. This is at variance with various empirical estimates
that give the opposite result, and that, in the $V$-band, vary between
$-0.24 \pm 0.16$ (Kennicutt et al. 1998) and about $-0.4$ mag/dex
(Kochanek 1997, Sasselov et al. 1997, Storm et
al. 1999\footnote{Recently, Storm (2000) suggested that this result
may not be confirmed from his latest analysis and that the correction
may have a positive sign instead.}). A mean of these four
determinations is $-0.38 \pm 0.09$ mag/dex, which for a difference in
metallicity of 0.4 dex, implies $\Delta M = -0.15 \pm 0.04$ in the
$V$-band. In the $I$-band we assume the same value following Sasselov
et al. (1997) and Kochanek (1997), but the reader should realise that
this number is less well established than the correction in $V$, and
in the $K$-band adopt $\Delta M = -0.07$.  Note however, that a
metallicity dependence as large as 0.4 mag in $V$ as suggested by
Sekiguchi \& Fukugita (1998) can be excluded at the 6$\sigma$ level
(Laney 1999, 2000). In recent papers, Saio \& Gautschy (1998) and
Sandage et al. (1999) found no significant metallicity dependence on
the bolometric $PL$-relation, and slopes of $-0.08$ mag/dex in $V$ and
$-0.1$ mag/dex in $I$ (Sandage et al. 1999), which represents a
shallower dependence than the values listed above, that are adopted in
the present study, and which therefore may be an extreme view.\\

\subsection{The distance to the LMC}

In Table~8 are listed the true distance moduli (DM) to the LMC for
$V,I,K$, the three slopes (`traditional' meaning the observed slopes
for the \C\ in the LMC, `shallower' adopting the theoretical slopes
from Bono et al. and `steeper' adopting the observed slopes for
Galactic \C\ from Gieren et al. (1998) and the three metallicity
corrections (`0' means no correction, `+' means a longer distance
scale as implied by the models from Bono et al., and `$-$' means a
shorter distance scale as implied from empirical evidence). The error
quoted includes the error in the zero point of both the Galactic and
LMC Cepheids, and where appropriate, the error due to the metallicity
correction, and for the solutions with either steeper or shallower
slope, the uncertainty due to the difference in DM at $\log P = 0.47$
and 0.77.

Also included are the weighted mean DM, averaged over $V,I,K$ (with
internal error), and, for reference, the (unweighted) mean DM per
photometric band of all the solutions (with the one-sided range in the
solutions).  Also included is the solution based on the
Wesenheit-index assuming the slope observed in the LMC, and no
metallicity correction. The DM range from 18.45 $\pm$ 0.18 to 18.86
$\pm$ 0.12. Several important conclusions may be drawn:

\bigskip

{\bf (1)} For every combination, the $PL$-relation in $K$ gives the
shortest distance, and the difference between the distance based on
$V$ and $K$ can be as large as 0.24 mag (solutions 4, 5 in Table~8). 


This systematic effect is worrying and merits investigation. It could
hint to errors in the reddening, or dereddening procedure. It is
illustrative to note that the (minimum, maximum, mean) extinction for
the whole sample is ($-0.21$, 3.4, 1.3) in $A_{\rm V}$, ($-0.08$, 2.1,
0.75) in $A_{\rm I}$, and ($-0.05$, 0.26, 0.10) in $A_{\rm K}$. This
implies that any uncertainty in reddening is less in $K$. In Sect.~5.1
we have estimated these uncertainties (about 0.04 in $V$, 0.03 in $I$
and 0.01 mag in $K$) and added them as a random errors. Possibly
these are errors of a systematic nature instead. It is interesting to
note that application of the procedures outlined in Sect.~3.2 results
in negative reddening in some cases.

One can raise the question how much bluer the period-colour-relations
need to be to give positive reddening for all stars. It turns out that
Eq.~(4) needs to be bluer by 0.065 mag, Eq.~(8) by 0.055 mag, and
Eq.~(11) by 0.075 mag. For the default slopes and using all stars, the
zero points in $V,I,K$ would change to, respectively, $-1.620$ (from
$-1.411$), $-1.970$ (from $-1.892$), and $-2.660$ (from $-2.607$).

On the other hand, the DM based on the Wesenheit-index, which avoids
the use of a $PC$-relation to estimate the individual reddenings, is
in perfect agreement with the DM based both on the $PL$-relations in $V$
and $I$. This suggests that the reddening is not the main reason for
the systematically shorter DM in the $K$-band. 

As pointed out in Sect.~4.3 there may be a bias in the $K$-band zero
point due to the smaller number of \C\ with accurate intensity-mean
magnitudes. The correction for this bias was estimated by comparing,
for the same sample of stars with $K$-band data, the zero point of the
$PL$-relations in $V$ and $I$ to those for the full samples in $V$ and
$I$, and this correction is about 0.1 mag, in the sense that it makes
the DM based on $K$ shorter than they would be without this
correction. This uncertainty can only be eliminated if more
intensity-mean NIR magnitudes come available.



{\bf (2)} The uncertainty in the type of metallicity correction introduces
a range in DM of up to 0.20 mag in $V,I$, and 0.12 mag in $K$.

{\bf (3)} The uncertainty in the slope of the Galactic $PL$-relations
introduces a range in DM of about 0.16 mag in $V,I$, and about 0.05
mag in $K$.\\


\noindent
Taking the case with the observed slopes of LMC Cepheids with no
metallicity correction as default, one may summarise the results as
follows.  Based on the $PL$-relation in $V$ and $I$, and the
Wesenheit-index, the true DM to the LMC is 18.60 $\pm$ 0.11 ($\pm$
0.08 slope) ($^{+0.08}_{-0.15}$ metallicity). Based on the
$PL$-relation in $K$ it is 18.52 $\pm$ 0.18 ($\pm$ 0.03 slope) ($\pm
0.06$ metallicity) ($^{+0.10}_{-0}$ sample bias). The terms between
parenthesis indicate the possible systematic uncertainties due to the
slope of the Galactic $PL$-relations, the metallicity corrections, and
in the $K$-band, due to the limited number of stars. Recent work by
Sandage et al. (1999) indicate that the effect of metallicity towards
shorter distances may be smaller in $V$ and $I$ than indicated here.
A more accurate determination is not possible without more definite
information on the slope of the Galactic $PL$-relations and the
metallicity correction. \\

Our prefered distance modulus is the one based on the $PL$-relation in
$V$, $I$ and the Wesenheit index, and puts the LMC 0.10 mag in DM
closer than the value of 18.70 derived by FC. The difference is due to
four effects that all work in the same direction, namely, (1) FC apply
a metallicty correction of +0.042 mag, (2) the difference in the zero
point in the Galactic $PL$-relation in $V$ between FC and this study
is +0.03 mag (+0.01 mag is due to Malmquist bias which FC did not take
into account, while +0.02 mag is due to the different sample and
slightly different photometry in some cases), (3) the difference in
the DM based on $V$ compared to the mean of the DM based on $V$, $I$
and the Wesenheit-index is +0.02 mag, and (4) the difference between
FC and this study in the adopted zero point of the $PL$-relation in
$V$ of the LMC Cepheids is +0.01 mag. \\

We finally note that the influence of the choice of slope and the
metallicity correction are (predicted to be) smallest in the $K$-band
as well as the uncertainty in the extinction correction. If the 20-30
closest \C\ without published NIR photometry could have their NIR
intensity-mean magnitudes determined, then the uncertainty due to the
small number of stars could be eliminated and the zero point in $K$
could be determined with an error that is a factor of two smaller than
is possible at present. \\

\begin{table}
\caption[]{Metallicity dependence of the absolute magnitude between
Galaxy and LMC from recent theoretical models. }
\begin{tabular}{rrrrl} \hline
$\log P_0$ & ${\Delta M}_{\rm V}$ & ${\Delta M}_{\rm I}$ & ${\Delta M}_{\rm K}$
& Reference \\ \hline
0.77 &  0.139 & 0.145 & 0.073 & Bono et al. (1999)\\
0.77 &  0.005 &$-0.007$ & 0.004 & Alibert et al. (1999)\\
0.47 & $-0.032$ & 0.062 & 0.025 & Bono et al. (1999)\\
0.47 & $-0.008$ &$-0.018$ &$-0.003$ & Alibert et al. (1999)\\
\hline
\end{tabular}
\end{table}

\begin{table*}
\caption[]{Distance Moduli to the LMC. Based on the $PL$-relations in
$V,I,K$ and the Wesenheit-index, for different assumptions about the
slope of the Galactic $PL$-relation, and metallicity correction.}
\begin{tabular}{crccccccc} \hline
Solution & Slope & Metallicity & $V$ & $I$ & $K$ & $W$ & Mean\\
         &       & dependence  &     &     &     &   & over $V,I,K$  \\ \hline
1& traditional & 0 & 18.62 $\pm$ 0.11 & 18.60 $\pm$ 0.12 & 18.52 $\pm$ 0.18
 & 18.60 $\pm$ 0.11 & 18.60 $\pm$ 0.07 & \\
2& traditional & + & 18.71 $\pm$ 0.12 & 18.70 $\pm$ 0.13 & 18.57 $\pm$ 0.18
 &         & 18.68 $\pm$ 0.08 & \\ 
3& traditional & $-$ & 18.47 $\pm$ 0.12 & 18.45 $\pm$ 0.12 & 18.45 $\pm$ 0.18
 &         & 18.46 $\pm$ 0.08 & \\
4& shallower & 0 & 18.80 $\pm$ 0.14 & 18.76 $\pm$ 0.19 & 18.57 $\pm$ 0.19
 &         & 18.73 $\pm$ 0.10 & \\
5& shallower & + & 18.86 $\pm$ 0.12 & 18.86 $\pm$ 0.16 & 18.62 $\pm$ 0.19
 &         & 18.81 $\pm$ 0.09 & \\
6& shallower & $-$ & 18.65 $\pm$ 0.14 & 18.61 $\pm$ 0.18 & 18.50 $\pm$ 0.19
 &        & 18.60 $\pm$ 0.10 & \\
7& steeper   & 0 & 18.66 $\pm$ 0.13 & 18.64 $\pm$ 0.15 & 18.53 $\pm$ 0.19
 &        & 18.63 $\pm$ 0.09 & \\
8& steeper   & + & 18.72 $\pm$ 0.18 & 18.75 $\pm$ 0.16 & 18.58 $\pm$ 0.20
 &        & 18.70 $\pm$ 0.10 & \\
9& steeper   & $-$ & 18.51 $\pm$ 0.14 & 18.49 $\pm$ 0.15 & 18.46 $\pm$ 0.19
 &        & 18.49 $\pm$ 0.09 & \\ \hline
mean/range & & & 18.67 / 0.20 & 18.65 / 0.21 & 18.53 / 0.09 & & 18.63 \\ 
\hline
\end{tabular}
\end{table*}

\subsection*{Acknowledgements}
We thank Pascal Fouqu\'e, Michael Feast and Patricia Whitelock for
providing tabular material in Gieren et al. (1998), respectively Feast
\& Whitelock (1997) in electronic format.  We thank Giuseppe Bono and
Santi Cassisi in calculating and communicating additional
$PL(C)$-relations to us.  Frederic Pont and Frederic Arenou are
thanked for lively and interesting discussions.  This research has
made use of the SIMBAD database, operated at CDS, Strasbourg, France.

\section*{Appendix}

In this appendix we list the sample of 236 \C\ in the \HP\ catalog
considered in this study. Listed in Table~A1 are the HIP number and
variable star name, the parallax and error in the parallax from the
\HP\ catalog except for RY Sco and Y Lac (Falin \& Mignard 1999), the
intensity-mean $V$ and $B-V$ adopted, the $\log$ of the fundamental
period, and the assumed pulsation mode. The $V-I$ colours and $JHK$
photometry can be found in G99, as described in Sect.~2.

\setcounter{table}{0}
\renewcommand{\thetable}{A\arabic{table}}

\begin{table*}
\caption[]{The sample of \HP\ \C}


\begin{tabular}{rrrrrrrr} \hline
HIP    & Name    &  $\pi$ & ${\sigma}_{\pi}$ & $V$  & $(B-V)$ & $\log
       P_0$ & mode \\
       &         &   (mas) & (mas)   &        &      &       &   \\ \hline
  1162 & FM Cas    &   0.10 &   1.27 &  9.127 &0.989 & 0.764 & 0 \\
  1213 & SY Cas    &   2.73 &   1.49 &  9.868 &0.992 & 0.610 & 0 \\
  2347 & DL Cas    &   2.32 &   1.09 &  8.969 &1.154 & 0.903 & 0 \\
  3886 & XY Cas    &  -0.02 &   1.58 &  9.935 &1.147 & 0.653 & 0 \\
  5138 & VW Cas    &  -2.12 &   3.61 & 10.697 &1.245 & 0.778 & 0 \\
  5658 & UZ Cas    &   4.37 &   3.64 & 11.338 &1.110 & 0.629 & 0 \\
  5846 & BP Cas    &  -0.60 &   2.04 & 10.920 &1.550 & 0.798 & 0 \\
  7192 & V636 Cas  &   1.72 &   0.81 &  7.199 &1.365 & 0.923 & 0 \\
  7548 & RW Cas    &   0.69 &   1.68 &  9.217 &1.196 & 1.170 & 0 \\
  8312 & BY Cas    &  -0.85 &   3.25 & 10.366 &1.309 & 0.662 & 1 \\
  8614 & VV Cas    &  -4.78 &   4.18 & 10.724 &1.143 & 0.793 & 0 \\
  9928 & VX Per    &   1.08 &   1.48 &  9.312 &1.158 & 1.037 & 0 \\
 11174 & V440 Per  &   1.62 &   0.83 &  6.282 &0.873 & 1.039 & 1 \\
 11420 & SZ Cas    &   2.21 &   1.60 &  9.853 &1.419 & 1.135 & 0 \\
 11767 & $\alpha$ UMi   &   7.56 &   0.48 &  1.982 &0.598 & 0.754 & 1 \\
 12817 & DF Cas    &  -0.27 &   3.65 & 10.848 &1.181 & 0.584 & 0 \\
 13367 & SU Cas    &   2.31 &   0.58 &  5.970 &0.703 & 0.440 & 1 \\
 19978 & SX Per    &  -1.59 &   2.96 & 11.158 &1.155 & 0.632 & 0 \\
 20202 & AS Per    &   0.56 &   1.84 &  9.723 &1.302 & 0.697 & 0 \\
 21517 & SZ Tau    &   3.12 &   0.82 &  6.531 &0.844 & 0.651 & 1 \\
 22445 & SV Per    &  -3.32 &   1.54 &  9.020 &1.029 & 1.046 & 0 \\
 23210 & AN Aur    &  -1.19 &   2.34 & 10.455 &1.218 & 1.012 & 0 \\
 23360 & RX Aur    &   1.32 &   1.02 &  7.655 &1.009 & 1.065 & 0 \\
 23768 & CK Cam    &  -0.33 &   0.95 &  7.541 &0.990 & 0.518 & 0 \\
 24105 & BK Aur    &   0.47 &   1.38 &  9.427 &1.062 & 0.903 & 0 \\
 24281 & SY Aur    &   1.15 &   1.70 &  9.074 &1.000 & 1.006 & 0 \\
 24500 & YZ Aur    &   3.70 &   2.10 & 10.332 &1.375 & 1.260 & 0 \\
 25642 & Y Aur     &  -0.40 &   1.47 &  9.607 &0.911 & 0.586 & 0 \\
 26069 & $\beta$ Dor  &   3.14 &   0.59 &  3.731 &0.807 & 0.993 & 0 \\
 27119 & ST Tau    &   3.15 &   1.17 &  8.217 &0.847 & 0.606 & 0 \\
 27183 & EU Tau    &   0.86 &   1.38 &  8.093 &0.664 & 0.473 & 1 \\
 28625 & RZ Gem    &   1.90 &   1.97 & 10.007 &1.025 & 0.743 & 0 \\
 28945 & AA Gem    &  -2.25 &   2.42 &  9.721 &1.061 & 1.053 & 0 \\
 29022 & CS Ori    &  -0.54 &   3.36 & 11.381 &0.924 & 0.590 & 0 \\
 29386 & GQ Ori    &   4.77 &   1.13 &  8.965 &0.976 & 0.935 & 0 \\
 30219 & SV Mon    &  -1.18 &   1.14 &  8.219 &1.048 & 1.183 & 0 \\
 30286 & RS Ori    &   2.02 &   1.45 &  8.412 &0.945 & 0.879 & 0 \\
 30827 & RT Aur    &   2.09 &   0.89 &  5.446 &0.595 & 0.572 & 0 \\
 31306 & DX Gem    &  -2.58 &   2.49 & 10.746 &0.936 & 0.650 & 1 \\
 31404 & W Gem     &   0.86 &   1.16 &  6.950 &0.889 & 0.898 & 0 \\
 31624 & CV Mon    &   3.76 &   2.77 & 10.299 &1.297 & 0.731 & 0 \\
 31905 & BE Mon    &  -0.28 &   2.12 & 10.578 &1.134 & 0.432 & 0 \\
 32180 & AD Gem    &  -0.18 &   1.60 &  9.857 &0.694 & 0.578 & 0 \\
 32516 & V508 Mon  &  -2.42 &   2.28 & 10.518 &0.898 & 0.616 & 0 \\
 32854 & TX Mon    &   0.00 &   2.47 & 10.960 &1.096 & 0.940 & 0 \\
 33014 & EK Mon    &  -0.77 &   2.69 & 11.048 &1.195 & 0.598 & 0 \\
 33520 & TZ Mon    &   1.61 &   2.12 & 10.761 &1.116 & 0.871 & 0 \\
 33791 & AC Mon    &   0.90 &   1.94 & 10.067 &1.165 & 0.904 & 0 \\
 33874 & V526 Mon  &   3.43 &   1.12 &  8.597 &0.593 & 0.579 & 1 \\
 34088 & $\zeta$ Gem   &   2.79 &   0.81 &  3.918 &0.798 & 1.007 & 0 \\
 34421 & V465 Mon  &   2.28 &   1.88 & 10.379 &0.762 & 0.586 & 1 \\
 34527 & TV CMa    &   0.90 &   1.97 & 10.582 &1.175 & 0.669 & 0 \\
 34895 & RW CMa    &   3.12 &   2.16 & 11.096 &1.225 & 0.758 & 0 \\
 35212 & RY CMa    &   0.96 &   1.09 &  8.110 &0.847 & 0.670 & 0 \\
 35665 & RZ CMa    &  -1.95 &   1.51 &  9.697 &1.004 & 0.629 & 0 \\
 35708 & TW CMa    &   1.26 &   1.51 &  9.561 &0.970 & 0.845 & 0 \\
 36125 & VZ CMa    &   1.58 &   1.65 &  9.383 &0.957 & 0.495 & 0 \\
 36617 & VW Pup    &  -5.65 &   2.83 & 11.365 &1.065 & 0.632 & 0 \\
 36685 & X Pup     &  -0.05 &   1.10 &  8.460 &1.127 & 1.414 & 0 \\
 37174 & MY Pup    &   0.65 &   0.52 &  5.677 &0.631 & 0.913 & 1 \\
\hline
\end{tabular}
\end{table*}

\setcounter{table}{0}
\begin{table*}
\caption[]{Continued}


\begin{tabular}{rrrrrrrr} \hline
HIP    & Name    &  $\pi$ & ${\sigma}_{\pi}$ & $V$  & $(B-V)$ & $\log
       P_0$ & mode \\
       &         &   (mas) & (mas)   &        &      &       &   \\ \hline
 37207 & VZ Pup    &   1.49 &   1.47 &  9.621 &1.162 & 1.365 & 0 \\
 37506 & EK Pup    &   3.54 &   2.34 & 10.664 &0.816 & 0.571 & 1 \\
 37511 & WW Pup    &   2.07 &   1.91 & 10.554 &0.874 & 0.742 & 0 \\
 37515 & WX Pup    &  -1.05 &   1.08 &  9.063 &0.968 & 0.951 & 0 \\
 38063 & AD Pup    &  -4.05 &   1.74 &  9.863 &1.049 & 1.133 & 0 \\
 38907 & AP Pup    &   1.07 &   0.64 &  7.371 &0.838 & 0.706 & 0 \\
 38944 & WY Pup    &   0.11 &   2.09 & 10.569 &0.791 & 0.720 & 0 \\
 38965 & AQ Pup    &   8.85 &   4.03 &  8.669 &1.337 & 1.479 & 0 \\
 39010 & LS Pup    &   3.90 &   2.71 & 10.442 &1.231 & 1.151 & 0 \\
 39144 & WZ Pup    &  -0.55 &   1.77 & 10.326 &0.789 & 0.701 & 0 \\
 39666 & BN Pup    &   4.88 &   1.72 &  9.882 &1.186 & 1.136 & 0 \\
 39840 & LX Pup    &   3.08 &   2.59 & 10.630 &1.032 & 1.142 & 0 \\
 40078 & HL Pup    & -10.42 &   4.30 & 10.740 &0.862 & 0.542 & 0 \\
 40155 & AH Vel    &   2.23 &   0.55 &  5.695 &0.579 & 0.782 & 1 \\
 40178 & AT Pup    &   1.20 &   0.74 &  7.957 &0.783 & 0.824 & 0 \\
 40233 & RS Pup    &   0.49 &   0.68 &  7.028 &1.434 & 1.617 & 0 \\
 41588 & V Car     &   0.34 &   0.58 &  7.362 &0.872 & 0.826 & 0 \\
 42257 & RZ Vel    &   1.35 &   0.63 &  7.079 &1.120 & 1.310 & 0 \\
 42321 & T Vel     &   0.48 &   0.72 &  8.024 &0.922 & 0.666 & 0 \\
 42831 & SW Vel    &   1.30 &   0.90 &  8.120 &1.162 & 1.370 & 0 \\
 42926 & SX Vel    &   1.54 &   0.79 &  8.251 &0.888 & 0.980 & 0 \\
 42929 & ST Vel    &  -1.62 &   0.99 &  9.704 &1.195 & 0.768 & 0 \\
 44847 & BG Vel    &   1.33 &   0.65 &  7.635 &1.175 & 0.999 & 1 \\
 45570 & DK Vel    &  -1.70 &   3.78 & 10.614 &0.774 & 0.546 & 1 \\
 45949 & W Car     &   1.16 &   0.63 &  7.589 &0.788 & 0.641 & 0 \\
 46746 & DR Vel    &  -0.45 &   1.07 &  9.520 &1.518 & 1.049 & 0 \\
 47177 & AE Vel    &  -0.64 &   1.33 & 10.262 &1.243 & 0.853 & 0 \\
 47854 & $l$ Car     &   2.16 &   0.47 &  3.724 &1.299 & 1.551 & 0 \\
 48122 & FN Vel    &   0.77 &   1.63 & 10.430 &0.912 & 0.726 & 0 \\
 48663 & GX Car    &   1.43 &   1.12 &  9.364 &1.043 & 0.857 & 0 \\
 50244 & CN Car    &   5.11 &   1.53 & 10.700 &1.089 & 0.693 & 0 \\
 50655 & RY Vel    &  -1.15 &   0.83 &  8.397 &1.352 & 1.449 & 0 \\
 50722 & AQ Car    &   1.02 &   0.81 &  8.851 &0.928 & 0.990 & 0 \\
 51142 & UW Car    &  -0.64 &   1.12 &  9.426 &0.971 & 0.728 & 0 \\
 51262 & YZ Car    &   1.79 &   1.03 &  8.714 &1.124 & 1.259 & 0 \\
 51338 & UX Car    &   0.00 &   0.87 &  8.308 &0.627 & 0.566 & 0 \\
 51894 & XX Vel    &   1.14 &   1.50 & 10.654 &1.162 & 0.844 & 0 \\
 51909 & UZ Car    &  -0.70 &   1.00 &  9.323 &0.875 & 0.716 & 0 \\
 52157 & HW Car    &  -0.71 &   1.06 &  9.163 &1.055 & 0.964 & 0 \\
 52380 & EY Car    &   3.46 &   1.62 & 10.318 &0.854 & 0.459 & 0 \\
 52570 & SV Vel    &  -1.27 &   0.97 &  8.524 &1.054 & 1.149 & 0 \\
 52661 & SX Car    &   2.48 &   1.06 &  9.089 &0.887 & 0.687 & 0 \\
 53083 & WW Car    &   4.23 &   1.39 &  9.743 &0.890 & 0.670 & 0 \\
 53397 & WZ Car    &  -0.41 &   1.14 &  9.247 &1.142 & 1.362 & 0 \\
 53536 & XX Car    &  -0.63 &   0.95 &  9.322 &1.054 & 1.196 & 0 \\
 53589 & U Car     &  -0.04 &   0.62 &  6.288 &1.183 & 1.589 & 0 \\
 53593 & CY Car    &  -0.30 &   1.40 &  9.782 &0.953 & 0.630 & 0 \\
 53867 & FN Car    &  -1.91 &   2.48 & 11.542 &1.101 & 0.661 & 0 \\
 53945 & XY Car    &  -0.62 &   0.95 &  9.295 &1.214 & 1.095 & 0 \\
 54101 & XZ Car    &  -0.30 &   0.96 &  8.601 &1.266 & 1.221 & 0 \\
 54543 & ER Car    &   1.36 &   0.69 &  6.824 &0.867 & 0.887 & 0 \\
 54621 & GH Car    &   0.43 &   1.03 &  9.177 &0.932 & 0.915 & 1 \\
 54659 & V898 Cen  &  -0.32 &   0.73 &  8.000 &0.574 & 0.547 & 0 \\
 54715 & IT Car    &   1.00 &   0.82 &  8.097 &0.990 & 0.877 & 0 \\
 54862 & GI Car    &  -0.41 &   1.10 &  8.323 &0.739 & 0.802 & 1 \\
 54891 & FR Car    &   0.35 &   1.29 &  9.661 &1.121 & 1.030 & 0 \\
 55726 & AY Cen    &  -0.24 &   1.04 &  8.830 &1.009 & 0.725 & 0 \\
 55736 & AZ Cen    &  -0.20 &   1.04 &  8.636 &0.653 & 0.660 & 1 \\
 56176 & V419 Cen  &   1.72 &   0.93 &  8.186 &0.758 & 0.898 & 1 \\
\hline
\end{tabular}
\end{table*}

\setcounter{table}{0}
\begin{table*}
\caption[]{Continued}


\begin{tabular}{rrrrrrrr} \hline
HIP    & Name    &  $\pi$ & ${\sigma}_{\pi}$ & $V$  & $(B-V)$ & $\log
       P_0$ & mode \\
       &         &   (mas) & (mas)   &        &      &       &   \\ \hline
 57130 & KK Cen    &  -1.84 &   2.89 & 11.480 &1.282 & 1.086 & 0 \\
 57260 & RT Mus    &   1.13 &   0.99 &  9.022 &0.834 & 0.489 & 0 \\
 57884 & BB Cen    &   2.85 &   1.27 &  9.781 &1.150 & 1.066 & 0 \\
 57978 & UU Mus    &   3.03 &   1.43 & 10.073 &0.953 & 0.757 & 1 \\
 59575 & AD Cru    &   1.87 &   2.32 & 11.051 &1.279 & 0.806 & 0 \\
 60259 & T Cru     &   0.86 &   0.62 &  6.566 &0.922 & 0.828 & 0 \\
 60455 & R Cru     &   1.97 &   0.82 &  6.766 &0.772 & 0.765 & 0 \\
 61136 & BG Cru    &   1.94 &   0.57 &  5.487 &0.606 & 0.678 & 1 \\
 61981 & R Mus     &   1.69 &   0.59 &  6.298 &0.757 & 0.876 & 0 \\
 62986 & S Cru     &   1.34 &   0.71 &  6.600 &0.761 & 0.671 & 0 \\
 63693 & V496 Cen  &   1.61 &   1.53 &  9.966 &1.172 & 0.646 & 0 \\
 64969 & V378 Cen  &   0.96 &   1.02 &  8.460 &1.035 & 0.969 & 1 \\
 65970 & V659 Cen  &   0.75 &   1.28 &  6.598 &0.758 & 0.750 & 0 \\
 66189 & VW Cen    &  -2.02 &   3.63 & 10.245 &1.345 & 1.177 & 0 \\
 66383 & KN Cen    &  -1.38 &   2.82 &  9.870 &1.582 & 1.532 & 0 \\
 66696 & XX Cen    &   2.04 &   0.94 &  7.818 &0.983 & 1.039 & 0 \\
 67566 & V381 Cen  &   1.13 &   0.91 &  7.653 &0.792 & 0.706 & 0 \\
 70203 & V339 Cen  &   0.33 &   1.16 &  8.753 &1.191 & 0.976 & 0 \\
 71116 & V Cen     &   0.05 &   0.82 &  6.836 &0.875 & 0.740 & 0 \\
 71492 & V737 Cen  &   3.71 &   0.84 &  6.719 &0.999 & 0.849 & 0 \\
 72264 & BP Cir    &   0.13 &   0.88 &  7.560 &0.702 & 0.531 & 1 \\
 72583 & AV Cir    &   3.40 &   1.09 &  7.439 &0.910 & 0.640 & 1 \\
 74448 & IQ Nor    &  -0.24 &   3.08 &  9.566 &1.314 & 0.916 & 0 \\
 75018 & R TrA     &   0.43 &   0.71 &  6.660 &0.722 & 0.530 & 0 \\
 75430 & GH Lup    &   2.65 &   0.86 &  7.635 &1.210 & 0.967 & 0 \\
 76918 & U Nor     &   2.52 &   1.28 &  9.238 &1.576 & 1.102 & 0 \\
 78476 & S TrA     &   1.59 &   0.72 &  6.397 &0.752 & 0.801 & 0 \\
 78797 & RS Nor    &  -0.23 &   1.81 & 10.027 &1.287 & 0.792 & 0 \\
 79625 & GU Nor    &   4.45 &   2.06 & 10.411 &1.273 & 0.538 & 0 \\
 79932 & S Nor     &   1.19 &   0.75 &  6.426 &0.945 & 0.989 & 0 \\
 82023 & V340 Ara  &   0.06 &   2.12 & 10.164 &1.539 & 1.318 & 0 \\
 82498 & KQ Sco    &   0.07 &   2.31 &  9.807 &1.934 & 1.458 & 0 \\
 83059 & RV Sco    &   2.54 &   1.13 &  7.040 &0.955 & 0.783 & 0 \\
 83674 & BF Oph    &   1.17 &   1.01 &  7.337 &0.868 & 0.609 & 0 \\
 85035 & V636 Sco  &  -0.45 &   0.89 &  6.654 &0.936 & 0.832 & 0 \\
 85701 & V482 Sco  &  -0.45 &   1.16 &  7.965 &0.975 & 0.656 & 0 \\
 86269 & V950 Sco  &   2.46 &   1.04 &  7.302 &0.775 & 0.683 & 1 \\
 87072 & X Sgr     &   3.03 &   0.94 &  4.549 &0.739 & 0.846 & 0 \\
 87173 & V500 Sco  &   2.21 &   1.30 &  8.729 &1.276 & 0.969 & 0 \\
 87345 & RY Sco    &   0.96 &   2.70 &  8.004 &1.426 & 1.308 & 0 \\
 87495 & Y Oph     &   1.14 &   0.80 &  6.169 &1.377 & 1.234 & 0 \\
 89013 & CR Ser    &  -3.04 &   2.08 & 10.842 &1.644 & 0.724 & 0 \\
 89276 & AP Sgr    &  -0.95 &   0.92 &  6.955 &0.807 & 0.704 & 0 \\
 89596 & WZ Sgr    &  -0.75 &   1.76 &  8.030 &1.392 & 1.339 & 0 \\
 89968 & Y Sgr     &   2.52 &   0.93 &  5.744 &0.856 & 0.761 & 0 \\
 90110 & AY Sgr    &  -0.99 &   2.28 & 10.549 &1.457 & 0.817 & 0 \\
 90241 & XX Sgr    &   2.64 &   1.22 &  8.852 &1.107 & 0.808 & 0 \\
 90791 & X Sct     &   0.97 &   1.46 & 10.006 &1.140 & 0.623 & 0 \\
 90836 & U Sgr     &   0.27 &   0.92 &  6.695 &1.087 & 0.829 & 0 \\
 91239 & EV Sct    &   0.91 &   1.92 & 10.137 &1.160 & 0.643 & 1 \\
 91366 & Y Sct     &   0.00 &   1.69 &  9.628 &1.539 & 1.015 & 0 \\
 91613 & CK Sct    &   3.62 &   2.12 & 10.590 &1.566 & 0.870 & 0 \\
 91697 & RU Sct    &   0.89 &   1.61 &  9.466 &1.645 & 1.294 & 0 \\
 91706 & TY Sct    &   4.02 &   2.27 & 10.831 &1.657 & 1.043 & 0 \\
 91738 & CM Sct    &  -3.72 &   2.35 & 11.106 &1.371 & 0.593 & 0 \\
 91785 & Z Sct     &   1.14 &   1.66 &  9.600 &1.330 & 1.111 & 0 \\
 91867 & SS Sct    &  -1.07 &   1.17 &  8.211 &0.944 & 0.565 & 0 \\
 92370 & YZ Sgr    &   0.87 &   1.03 &  7.358 &1.032 & 0.980 & 0 \\
 92491 & BB Sgr    &   0.61 &   0.99 &  6.947 &0.987 & 0.822 & 0 \\
\hline
\end{tabular}
\end{table*}

\setcounter{table}{0}
\begin{table*}
\caption[]{Continued}


\begin{tabular}{rrrrrrrr} \hline
HIP    & Name    &  $\pi$ & ${\sigma}_{\pi}$ & $V$  & $(B-V)$ & $\log
       P_0$ & mode \\
       &         &   (mas) & (mas)   &        &      &       &   \\ \hline
 93063 & V493 Aql  &  -2.77 &   2.43 & 11.083 &1.280 & 0.475 & 0 \\
 93124 & FF Aql    &   1.32 &   0.72 &  5.372 &0.756 & 0.806 & 1 \\
 93399 & V336 Aql  &   0.75 &   1.47 &  9.848 &1.312 & 0.864 & 0 \\
 93681 & SZ Aql    &   0.20 &   1.10 &  8.599 &1.389 & 1.234 & 0 \\
 93990 & TT Aql    &   0.41 &   0.96 &  7.141 &1.292 & 1.138 & 0 \\
 94004 & V496 Aql  &  -3.81 &   1.05 &  7.751 &1.146 & 0.833 & 0 \\
 94094 & FM Aql    &   2.45 &   1.11 &  8.270 &1.277 & 0.786 & 0 \\
 94402 & FN Aql    &   1.53 &   1.18 &  8.382 &1.214 & 1.138 & 1 \\
 94685 & V473 Lyr  &   1.94 &   0.62 &  6.182 &0.632 & 0.433 & 2 \\
 95118 & V600 Aql  &   1.42 &   1.80 & 10.037 &1.462 & 0.860 & 0 \\
 96458 & U Vul     &   0.59 &   0.77 &  7.128 &1.275 & 0.903 & 0 \\
 96596 & V924 Cyg  &   0.83 &   1.64 & 10.710 &0.847 & 0.746 & 0 \\
 97309 & BR Vul    &  -2.80 &   1.70 & 10.687 &1.474 & 0.716 & 0 \\
 97439 & V1154 Cyg &   0.88 &   0.88 &  9.190 &0.925 & 0.692 & 0 \\
 97717 & SV Vul    &   0.79 &   0.74 &  7.220 &1.442 & 1.653 & 0 \\
 97794 & V1162 Aql &   0.15 &   1.15 &  7.798 &0.879 & 0.888 & 1 \\
 97804 & $\eta$ Aql   &   2.78 &   0.91 &  3.897 &0.789 & 0.856 & 0 \\
 98085 & S Sge     &   0.76 &   0.73 &  5.622 &0.805 & 0.923 & 0 \\
 98212 & X Vul     &  -0.33 &   1.10 &  8.849 &1.389 & 0.801 & 0 \\
 98376 & GH Cyg    &   1.93 &   1.67 &  9.924 &1.266 & 0.893 & 0 \\
 98852 & CD Cyg    &   0.46 &   1.00 &  8.947 &1.266 & 1.232 & 0 \\
 99276 & V402 Cyg  &   1.19 &   1.18 &  9.873 &1.008 & 0.640 & 0 \\
 99567 & MW Cyg    &  -1.63 &   1.30 &  9.489 &1.316 & 0.775 & 0 \\
 99887 & V495 Cyg  &  -0.95 &   1.32 & 10.621 &1.623 & 0.827 & 0 \\
101393 & SZ Cyg    &   0.86 &   1.09 &  9.432 &1.477 & 1.179 & 0 \\
102276 & X Cyg     &   1.47 &   0.72 &  6.391 &1.130 & 1.214 & 0 \\
102949 & T Vul     &   1.95 &   0.60 &  5.754 &0.635 & 0.647 & 0 \\
103241 & V520 Cyg  &   1.51 &   1.73 & 10.851 &1.349 & 0.607 & 0 \\
103433 & VX Cyg    &   0.88 &   1.43 & 10.069 &1.704 & 1.304 & 0 \\
103656 & TX Cyg    &   0.50 &   1.09 &  9.511 &1.784 & 1.168 & 0 \\
104002 & VY Cyg    &  -0.02 &   1.44 &  9.593 &1.215 & 0.895 & 0 \\
104185 & DT Cyg    &   1.72 &   0.62 &  5.774 &0.538 & 0.549 & 1 \\
104564 & V459 Cyg  &   0.51 &   1.50 & 10.601 &1.439 & 0.860 & 0 \\
104877 & V386 Cyg  &   2.22 &   1.17 &  9.635 &1.491 & 0.721 & 0 \\
105369 & V532 Cyg  &   0.84 &   0.94 &  9.086 &1.036 & 0.516 & 0 \\
106754 & V538 Cyg  &   0.10 &   1.52 & 10.456 &1.283 & 0.787 & 0 \\
107899 & VZ Cyg    &   2.84 &   1.17 &  8.959 &0.876 & 0.687 & 0 \\
108426 & IR Cep    &   1.38 &   0.61 &  7.784 &0.888 & 0.476 & 1 \\
108427 & CP Cep    &   1.54 &   1.52 & 10.590 &1.668 & 1.252 & 0 \\
108630 & BG Lac    &  -0.35 &   1.31 &  8.883 &0.949 & 0.727 & 0 \\
109340 & Y Lac     &   0.45 &   1.70 &  9.146 &0.731 & 0.636 & 0 \\
110964 & AK Cep    &   0.22 &   2.52 & 11.180 &1.341 & 0.859 & 0 \\
110968 & V411 Lac  &   0.67 &   0.70 &  7.860 &0.741 & 0.464 & 0 \\
110991 & $\delta$ Cep  &   3.32 &   0.58 &  3.954 &0.657 & 0.730 & 0 \\
111972 & Z Lac     &   2.04 &   0.89 &  8.415 &1.095 & 1.037 & 0 \\
112026 & RR Lac    &   0.94 &   0.95 &  8.848 &0.885 & 0.807 & 0 \\
112430 & CR Cep    &   1.67 &   1.06 &  9.656 &1.396 & 0.795 & 0 \\
112626 & V Lac     &   0.34 &   0.85 &  8.936 &0.873 & 0.697 & 0 \\
112675 & X Lac     &   0.57 &   0.79 &  8.407 &0.901 & 0.893 & 1 \\
114160 & SW Cas    &   1.07 &   1.37 &  9.705 &1.081 & 0.736 & 0 \\
115390 & CH Cas    &   0.21 &   1.68 & 10.973 &1.650 & 1.179 & 0 \\
115925 & CY Cas    &   2.76 &   3.21 & 11.641 &1.738 & 1.158 & 0 \\
116556 & RS Cas    &   2.43 &   1.24 &  9.932 &1.490 & 0.799 & 0 \\
116684 & DW Cas    &   1.19 &   1.95 & 11.112 &1.475 & 0.699 & 0 \\
117154 & CD Cas    &   1.91 &   1.58 & 10.738 &1.449 & 0.892 & 0 \\
117690 & RY Cas    &   0.02 &   1.38 &  9.927 &1.384 & 1.084 & 0 \\
118122 & DD Cas    &   0.57 &   1.14 &  9.876 &1.188 & 0.992 & 0 \\
118174 & CF Cas    &  -3.20 &   2.16 & 11.136 &1.174 & 0.688 & 0 \\
\hline
\end{tabular}
\end{table*}

{}

\end{document}